\documentclass[a4paper,12pt]{article}
\usepackage{amssymb,amsmath,mathrsfs}
\usepackage[usenames,dvipsnames]{xcolor}
\usepackage{hyperref, comment}
\usepackage{tensor}
\usepackage{graphicx}
\usepackage[left=1.2in,top=1in,right=1.2in,bottom=1in,headheight=0.8in,foot=0.5in]{geometry}
\usepackage[nodisplayskipstretch]{setspace} 
\usepackage[width=\textwidth]{caption}
\usepackage{pifont}% http://ctan.org/pkg/pifont
\usepackage{xcolor}
\usepackage{amsmath,latexsym}
\usepackage{outlines}
\usepackage{xcolor}
\usepackage[textsize=tiny]{todonotes}

\usepackage[indentafter]{titlesec}
\titleformat{name=\section}{}{\thetitle.}{0.8em}{\centering\scshape}
\titleformat{name=\subsection}[runin]{}{\thetitle.}{0.5em}{\bfseries}[.]
\titleformat{name=\subsubsection}[runin]{}{\thetitle.}{0.5em}{\itshape}[.]
\titleformat{name=\paragraph,numberless}[runin]{}{}{0em}{}[.]
\titlespacing{\paragraph}{0em}{0em}{0.5em}
\titleformat{name=\subparagraph,numberless}[runin]{}{}{0em}{}[.]
\titlespacing{\subparagraph}{0em}{0em}{0.5em}

\usepackage[style=ext-authoryear-comp,maxcitenames=2,uniquename=false,backend=biber,uniquelist=false,articlein=false]{biblatex}
\DeclareNameAlias{sortname}{last-first}

\newcommand{\nocontentsline}[3]{}
\newcommand{\tocless}[2]{\bgroup\let\addcontentsline=\nocontentsline#1{#2}\egroup}
\hypersetup{
	colorlinks=true,         
	linkcolor=Black,          
	citecolor=MidnightBlue,
	urlcolor=MidnightBlue            
 }

\newcommand{\beq}{\begin{eqnarray}}
\newcommand{\eeq}{\end{eqnarray}}

\title{Underdetermination in Classic and Modern Tests of General Relativity\vspace{-2em}}
\date{}

\makeatletter
\let\uppercasenonmath\@gobble

\begin{document}
\setstretch{1.0}
\maketitle

\begin{center}
\author{William J.~Wolf\footnote{Faculty of Philosophy, University of Oxford, UK. Email: william.wolf@philosophy.ox.ac.uk},~Marco Sanchioni\footnote{Università degli Studi di Urbino, Italy. Email: marco.sanchioni2@gmail.com},~ and James Read\footnote{Faculty of Philosophy, University of Oxford, UK. Email: james.read@philosophy.ox.ac.uk}}
\end{center}

\begin{abstract}
\noindent Canonically, `classic' tests of general relativity (GR) include perihelion precession, the bending of light around stars, and gravitational redshift; `modern' tests have to do with, \emph{inter alia}, relativistic time delay, equivalence principle tests, gravitational lensing, strong field gravity, and gravitational waves. The orthodoxy is that both classic and modern tests of GR afford experimental confirmation of that theory \emph{in particular}. In this article, we question this orthodoxy, by showing there are classes of both relativistic theories (with spatiotemporal geometrical properties different from those of GR) and non-relativistic theories (in which the lightcones of a relativistic spacetime are `widened') which would also pass such tests. Thus, (a) issues of underdetermination in the context of GR loom much larger than one might have thought, and (b) given this, one has to think more carefully about what exactly such tests in fact \emph{are} testing.
\end{abstract}
\tableofcontents
\setstretch{1.2}

%%this line gets rid of weird table of content spacing issue when you automatically skip lines between paragraphs
% \begingroup
% \makeatletter
% \parskip\z@skip
% \tableofcontents
% \endgroup

\section{Introduction}

%\noindent The history of gravitational research has featured some of the most spectacular and celebrated empirical successes within science. Newton's law of universal gravitation led to an unprecedented period of quantitative prediction, investigation, and discovery in celestial motion. Einstein's general theory of relativity (GR) then signalled a radical paradigm change, and in the process resolved many outstanding anomalies and opened up entirely new research avenues.

\noindent Einstein's general theory of relativity (GR) is our best theory of space and time, and has captured both the popular and the academic imagination. How could it not? The theory tells a compelling story, informing us not only that our most commonly held intuitions concerning space and time are wrong, but also that space and time themselves are inextricably interwoven, both together and with the material content of the world. According to GR, gravity is not \textit{really} a force, but rather a manifestation of spacetime curvature, which governs the motion of all matter in the universe.\footnote{We mean this in a loose sense---we wish to remain agnostic about which object in GR represents the gravitational field \emph{sensu stricto}. For more on this latter issue, see \parencite{Lehmkuhl2008}.}

Furthermore, the story goes, GR is one of the most well-confirmed theories in all of science as it has been subjected for over a century to rigorous tests and held up to the strictest scrutiny. Such testing and empirical success is understood to vindicate the theory in its entirety. But does this mean that all of GR's radical conclusions regarding the nature space, time, and gravity are thereby confirmed conclusively? In this article, we analyze closely the classic and modern tests of GR in order to explore the nature of these claims and examine what these tests in fact do manage to confirm conclusively---we find that the situation is more delicate than one might have thought.

To be specific, here are the two classes of tests of GR which we consider in this article:
\begin{description}
    \item[Classic tests] perihelion precession, bending of light, and gravitational redshift.
    \item[Modern tests] strong field gravity, gravitational lensing, FLRW cosmological solutions, gravitational waves, and black holes.
\end{description}

While this story of empirical success is certainly remarkable, we argue that the limitations of what these tests have actually shown are not widely appreciated. Indeed, while the extent to which these tests either disconfirm or severely constrain theories of gravity that are known to be \textit{empirically inequivalent} alternatives to GR is well-appreciated---this has been a major theme in modern gravitational research! (see e.g.~\textcite{Will})---what is not appreciated is that there are alternative theories that can be understood to be \textit{empirically equivalent} to GR in important ways and for which these tests offer far less in terms of conclusive analysis.\footnote{While we will not address this here, a separate concern that has been brought up in the context of empirical tests of gravitational physics is the theory-ladenness of such tests and observations. See \textcite{Elder2023-ELDTTI} for further discussion.} In particular, the classic tests of GR are significantly less constraining than is often appreciated---often, for example, they amount to tests of a particular geometric solution \emph{within} GR; a solution that can readily be reproduced by a number of alternative theories. Although modern tests are in a certain sense more stringent, the data resulting from these tests which is taken to confirm GR can likewise be accounted for by a number of alternative theories. In what follows in this article, we analyze how and to what extent both classes of test can be passed by theories of gravity other than GR; we do so by modifying systematically the geometrical content of GR (in particular, its projective and conformal structure) while retaining empirical predictions with respect to these tests. Thereby, we demonstrate that there are still persistent underdetermination issues even within the context of what we consider to be one of the most well-tested and rigorously-confirmed theories of physics.

The structure of the article is as follows. In \S\ref{s2}, we remind the reader of the issues in general philosophy of science relevant to this article---in particular, issues of underdetermination of theory by evidence and of theory equivalence. In \S\ref{s3}, we present the classic tests of GR. In \S\ref{s4}, we introduce the distinction between projective and conformal structures---these structures will be useful in later sections when it comes to classifying spacetime theories alternative to GR. In \S\ref{s5}, we consider the `geometric trinity' of relativistic alternatives to GR (reviewed by \textcite{Heisenberg}), and demonstrate that all theories in this `trinity' would pass both the classic and modern tests of GR. In \S\ref{s6}, we consider recently-developed non-relativistic spacetime theories (reviewed by \textcite{Obers}) which also pass the classic tests of GR, as well as (almost!) all the modern tests of GR. In \S\ref{s7}, we consider how to respond to the issues of underdetermination presented by these cases. In \S\ref{s8}, we wrap up.

%\textbf{to do later but once everything is done put in paragraph that breaks down paper by section}

\section{Underdetermination and Theory Equivalence}\label{s2}

\noindent In this section, we review issues of underdetermination of theory by evidence (\S\ref{s2.1}) and of theory equivalence (\S\ref{s2.2}).

\subsection{Weak and Strong Underdetermination}\label{s2.1}

Underdetermination of theory by evidence comes in different stripes. `Weak underdetermination' (also known as `transient underdetermination') refers to underdetermination with respect to the currently available data \parencite[p.~246]{Ladyman2001-LADUPO-2}. This occurs between distinct theories that give different predictions for at least some empirical phenomena. However, these predictions either have not yet been tested or cannot currently be tested, meaning that such underdetermination could conceivably be broken in the future. This is a familiar theme in current gravitational research, and in particular within cosmology. For example, the standard model of cosmology, dubbed the $\Lambda$CDM (`Lambda-cold dark matter') model, describes a universe completely consistent with the FLRW (Friedmann-Lemaitre-Robertson-Walker) solution of GR. Yet, puzzles concerning the nature of dark matter and dark energy have led cosmologists to explore a vast number of theories which explain $\Lambda$ and/or CDM, in ways that modify the typical understanding of these entities within a GR framework. A few (amongst many) examples include Brans-Dicke gravity, $f(R)$ gravity, quintessence, and certain relativistic extensions of modified Newtonian dynamics (MOND).\footnote{See \textcite{Clifton:2011jh, Joyce:2016vqv, Wolf:2023uno} for physics discussion on various modified gravity and dark energy proposals aimed at accounting for the current epoch of accelerating expansion and \textcite{DuerrWolf, Martens:2020lto} for some philosophical analysis on dark matter and modified gravity.} Another cosmological example can be found in examining attempts to model the early universe, where there are a number of theoretical frameworks in play such as inflation, bouncing cosmologies, and string gas cosmology.\footnote{See \textcite{Guth:2013sya, Ijjas:2015hcc, Brandenberger:2016vhg} for physics discussions of these theories and \textcite{DawidManuscript-DAWIIC, Wolf:2022yvd, WolfMeta, WolfDuerr} for philosophical analyses of some of the extra-empirical philosophical issues at play in these debates.} These theories can and do give rise to distinct predictions that differentiate them from each other, but current observational constraints are consistent with a number of possibilities. The hope is that future observations will break this underdetermination. 

In contrast to weak underdetermination, `strong underdetermination' refers to underdetermination with respect to all empirical data that will ever be available \parencite[pp.~261-262]{Ladyman2001-LADUPO-2}. This is often discussed in terms of distinct but empirically equivalent models within the same theory, where this refers usually to empirically equivalent models related by a symmetry. A famous example invites us to consider two Newtonian models of the universe, one at absolute rest and the other boosted by a constant velocity \parencite[pp.~46-47]{VanFraassenBas1980-VANTSI}. This example illustrates that there is a plurality of distinct Newtonian models of the universe which are compatible with the empirical data, yet an observer embedded in any of the worlds represented by those models could never distinguish empirically between them because the theory itself indicates that such absolute standards of motion are unobservable.\footnote{For some recent discussions as to whether absolute velocities really are empirically unobservable in Newtonian mechanics, see \textcite{SMR, Caspar}. Note also that one might still be able to distinguish between (some) empirically indistinguishable worlds using the linguistic resources of indexicals---for some discussion of this point, see \textcite{ChengRead}, but we won't go into this further in this article.} However, strong underdetermination need not be restricted to empirically equivalent models of a particular theory. Strong underdetermination can also exist between models of distinct yet empirically equivalent \emph{theories}. This has been considered in the recent philosophy literature in the context of theories related by dualities---see e.g.~\textcite{Butterfield, DeHaro2017-DEHASF-2, Matsubara2013-MATRUA, Read2016-REATIO-3}.

Strong underdetermination is understandably seen as a serious threat to scientific realism because the existence of empirically equivalent yet putatively distinct models calls into question the extent to which our scientific models can correspond fully to reality---surely, the realist thought goes, at most \emph{one} of those models can correspond to reality, so what gives? If multiple, ontologically distinct models can correctly describe the same phenomena, what license do we have for adopting a realist attitude towards the non-observable structures in the models of our theories? There is thus a serious motivation to deny that instances of strong underdetermination truly exist.

%Realist responses to strong underdetermination along these lines usually fall into one of three categories, all of which involve an interpretive move: (i) the discrimination approach, (ii) the common core approach, (iii) the overarching theory approach (Le Bihan/Read). (i) The discrimination approach, as the name suggests, seeks a principled reason to discriminate between the models and privilege the ontology claims of one of them. (ii) The common core approach seeks to isolate the `common core' that is shared between the distinct models or theories and then reinterpret this shared common core as representing the ontologically relevant information. (iii) The overarching approach seeks to embed the distinct models or theories into an overarching framework that allows one to reinterpret the distinct models or theories in light of a new ontology. (i) Breaks the underdetermination by privileging one of the theories over the other, while both (ii) and (iii) break the underdetermination by finding an interpretive framework that allows us to understand the models as one and the same. 

In this article, we argue that there are underappreciated instances of both strong and weak underdetermination within gravitational physics at the level of the most significant tests of gravity. For example, there are theories that are known to be dynamically equivalent to GR (i.e.,~they are equivalent at the level of their equations of motion---more on this in \S\ref{s5}), but which postulate very different kinds of geometrical structures in their description of gravity. These are the `Teleparallel Equivalent to General Relativity' (TEGR), which is by now somewhat known in the philosophical literature and describes gravitational effects using torsion in a flat spacetime, and the `Symmetric Teleparallel Equivalent to General Relativity' (STEGR), which is far less well-known in the existing philosophical literature and describes gravity as a manifestation of non-metricity in a flat spacetime.\footnote{Here, by `flat', we mean in the sense of affine geometry: the Riemann tensors associated to the affine connections of both TEGR and STEGR are vanishing.}

Crucially, the situation with these theories is different from the other myriad of relativistic gravitational theories that compete with GR (e.g.,~scalar-tensor, $f(R)$, vector-tensor, bimetric, etc.---see e.g.~\textcite{Clifton:2011jh}). Even though large numbers of these kinds of theories can pass the same experimental tests that GR does, they are inequivalent at the level of dynamics and have additional degrees of freedom that can be tuned to fit within empirical constraints; and thus represent a case of transient underdetermination that future observations will impact. As we shall see, no empirical test could ever discriminate between the theories in the trinity,\footnote{At least within the regime of classical physics: the theories do differ by boundary terms (see \textcite{WolfRead} for philosophical discussion) which might manifest as instantionic effects after path integral quantisation---this, however, deserves to be worked out in detail, and we won't discuss it further in this article, save for some brief remarks in \S\ref{s8}. It's also worth making the technical point here that in making this claim, we're restricting to GR models set on parallelizable manifolds: doing so undermines none of the points which we'll go on to make in this article.} which are collectively known as the `geometrical trinity' of relativistic gravitational theories \parencite{Heisenberg}.\footnote{It should also be noted that, as shown by \textcite{Wolf:2023rad} there also exists a `non-relativistic geometric trinity' of gravity that is closely analogous to the relativistic trinity considered here. See \textcite{Wolf:2023rad, March:2023trv} for further discussion. }\textsuperscript{,}\footnote{The theories within the geometric trinity are not the only theories one could in principle consider here as there are other theories that have significant overlap in their solution space with GR. For example, shape dynamics is another theory that can also be understood to be dynamically equivalent to GR in a sense (albeit with some caveats---see \textcite{Gomes:2012hq, Gomes:2011zi}). However, this theory is a far more radical departure from GR than TEGR or STEGR as it is based on entirely different symmetries (i.e.\ spatial diffeomorphisms and conformal Weyl symmetry---see \textcite{Barbour:2011dn}) and we will not consider it explicitly here.} Modulo an interpretive move to collapse this trinity into one theory, this is arguably a live case of strong underdetermination; one for which we will analyze and consider potential responses to in \S\ref{s7}. What we have in mind here is, then, akin to the `permanent' underdetermination identified by \textcite{PittsPermanent} in the case of massive spin-2 gravity; indeed, we regard our project here as being complementary to (but distinct from) that work.\footnote{See also Lorentz-violating massive gravity, as presented in e.g.~\textcite{Blas}.}
 
In addition to the geometric trinity, however, there is a case of \emph{weak} underdetermination in the context of the tests of GR which has to this point gone largely unnoticed within the literature. There is an even less familiar theory, known as `Type-II Newton Cartan Theory' (NCTII),\footnote{Partly this theory is not well known because it first appeared in the literature only in 2014---see \textcite{cite-key}; moreover, the action principle for the theory appeared in the literature only in 2019, with \textcite{ObersAction}.} which describes gravity in terms of a torsionful, fully non-relativistic spacetime \parencite{Obers}. While this theory differs from the theories in the geometric trinity in important ways that do end up distinguishing their empirical claims from each other (but---interestingly---only in the context of some very recent modern tests of GR, to do with gravitational wave physics, more on which below), this theory can also be understood to pass \emph{all} the classic tests of GR \parencite{ObersTests}. Furthermore, this theory does not merely pass these tests in the same way that, say, Brans-Dicke theory can be understood to pass solar system tests of GR (i.e.\ by tuning additional parameters to better approximate the GR result within experimental error). Rather, as we shall see, NCTII gives completely \textit{identical} predictions as GR for these important solar system tests and can be understood to be empirically equivalent to GR at the level of these tests. This leads to the bizarre conclusion that this instance of weak underdetermination between relativistic theories of gravity and non-relativistic theories of gravity has only been broken far more recently than most would imagine---indeed, with the advent of LIGO!

\subsection{Theory Equivalence}\label{s2.2}

As should be clear from the preceding discussion of underdetermination, it is important to establish what is meant by `equivalence' with regard to determining whether two theories are truly distinct from each other, or in fact are merely different formulations of the same theory. For example, the standard line goes that Heisenberg matrix mechanics and Schrödinger wave mechanics are not only empirically equivalent, but in fact are different formulations of the same theory of quantum mechanics.\footnote{Actually, it's questionable whether this claim is completely correct: see \textcite{Muller1997-MULTEM, Muller1997-MULTEM-2}. But the example is sufficient to illustrate our point.}
%The differences are understood to be cosmetic rather than substantial.\todo{This might not be the best example, in light of Muller's lovely 1997 papers in SHPMP. I think at the very least we should acknowledge Muller in a footnote here.}
How do we determine whether or not our gravitational theories introduced above are truly equivalent to one another? The literature concerning questions on theoretical equivalence is vast and offers many possible answers to this question \parencite{weatherall2019part, weatherall2019part2}. Of particular relevance to us are notions of `empirical equivalence' and `interpretational equivalence', to both of which we have already alluded.

%really clear that some notion of theory One go-to response to claims of strong underdetermination questions whether there are indeed any genuine examples of truly distinct yet empirically equivalent theories. Accordingly, the thought goes that either there are empirical differences which we are missing between the two theories (and thus the putative case of strong underdetermination is in fact a case of weak underdetermination) or what we perceive of as being different theories are in fact merely different formulations of the same theory. For example, matrix mechanics and wave mechanics are empirically equivalent versions of quantum mechanics, but are quite universally held to be different formulations of the same theory of quantum mechanics. The differences are understood to be cosmetic rather than substantial.\todo{This might not be the best example, in light of Muller's lovely 1997 papers in SHPMP. I think at the very least we should acknowledge Muller in a footnote here.} How do we determine whether these gravitational theories are truly equivalent to each other or not? The literature concerning questions on theoretical equivalence is vast and offers many possible answers to this question (Weatherall). Of particular relevance to us are notions of `empirical equivalence', to which we have already alluded, and `interpretational equivalence'.

Empirical equivalence is generally taken to be a necessary but not sufficient condition  for full theoretical equivalence. Complete empirical equivalence would mean that two theories have the same range of applicability regarding the empirical scenarios which they describe and provide indistinguishable predictions for said empirical scenarios. To be slightly more specific, we can understand (to use the terminology of \textcite{VanFraassenBas1980-VANTSI}) that models $M$ of a theory $T$ have `empirical substructures', which can represent observable phenomena. Suppose, for every $M$ of $T$, there is an $M'$ of $T'$, where the empirical substructures of $M$ and $M'$ are isomorphic, and vice versa. Then, $T$ and $T'$ can be understood to be empirically equivalent. 

This standard for empirical equivalence distinguishes between the theories within the geometric trinity on the one hand and NCTII on the other because (as we'll see in detail below) NCTII appears not to be able to support the same claims as the geometric trinity with regard to the phenomena of gravitational waves. Thus, we have here a case of \emph{weak} underdetermination that modern (but not classic!) tests of GR conceivably can break.

Reaching a conclusive verdict regarding the equivalence or inequivalence of the gravitational theories \emph{within} the geometric trinity, on the other hand, is a far more subtle business. These theories differ quite substantially in terms of the structures from which they are built, yet they can all be understood to be dynamically equivalent to each other as their actions differ only by boundary terms. With regard to dynamical content, these theories are equivalent full stop. As the empirical tests of GR that have been performed to date have only been sensitive to dynamical content, we here set aside questions concerning demonstration of equivalence at the level of empirical claims regarding boundary-related content and phenomena (these issues are discussed further by \textcite{WolfRead}).\footnote{Even if it does seem that claims of empirical equivalence can be supported regarding boundary phenomena (for example, both theories reproduce the GR result for black hole entropy as shown by \textcite{Oshita:2017nhn, Heisenberg:2022nvs}), it is important to note that this is a non-trivial matter that requires further investigation and is often ignored in the literature on theoretical equivalence.} For our purposes here, the geometric trinity clearly seems to constitute an example of strong underdetermination, where both classic and modern tests that have been performed to date offer no hope of discriminating amongst the theories.

However, could we still not just say that all of these theories within the trinity are actually somehow the same `theory', but merely dressed up in different mathematical details or formalisms? One way of doing this would be to demonstrate that these theories can be understood as interpretationally equivalent to each other. Interpretational equivalence in this sense holds when theories are understood to make all of the same claims about the phenomena they describe, going beyond purely empirical considerations \parencite{Coffey2014-COFTEA}.\footnote{We discuss interpretational equivalance further in \S\ref{s7}, in the context of some arguments made by \textcite{Knox2011-KNONTA}.} This would include claims about what kinds of entities exist in the world and what properties they have, as well as what the fundamental laws of nature are. If this can be done, this would offer an avenue towards breaking the underdetermination. However, arguing that the geometric trinity theories are interpretationally equivalent to each other in their current forms is not the only way to proceed. Indeed one could also move to new interpretive frameworks to cash out their equivalence. We return to this issue in \S\ref{s7}, but before doing so we need to say more regarding how all the theories introduced up to this point interact with the tests of GR, both classic and modern.

\section{Classic Tests of General Relativity}\label{s3}

\noindent The three classic tests of GR were all identified by Einstein early on in his development of the theory (see \textcite{Einstein:1916vd} for an early summary and discussion of these tests\footnote{See \textcite{Lehmkuhl} for an excellent recent discussion of Einstein on the classic tests of GR.}). Given their temporal proximity to  the development of GR, explanatory power, and novel nature, they were hailed as spectacular confirmations and christened a new paradigm for space, time, and gravitation. The classic tests are:
%The perihelion precession made use of data that was already known and had yet to be explained. The bending of light was a completely novel prediction and confirmed within a few years of the the theory's publication. The gravitational redshift of light had to wait a few decades, but completed the trifecta of Einstein's proposed tests. 
\begin{enumerate}
    \item Perihelion precession of Mercury's orbit: It had long been known that Mercury's perihelion had an anomalous precession of about 43 arcseconds per century \parencite{1859AnPar...5....1L, 1882USNAO...1..363N}. Orbital precession is predicted by Newtonian gravity in the presence of perturbing forces and the total precession was an order of magnitude larger than the unexplained value as most of it had been accounted for by determining the perturbing affects of other solar system bodies. However, the 43 arcseconds remained persistently unaccounted-for. Given that this empirical result was already known, explaining this was a `simple' matter of solving the Kepler two-body problem in GR and noticing that additional terms in the effective gravitational potential produce this effect. 
    \item Gravitational deflection of light: This is another effect that has a Newtonian analogue as it was understood that even in a Newtonian framework gravitational effects should bend the paths of light rays \parencite{2005physics...8030S}. However, Einstein correctly used GR to predict that the bending of light due to the Sun's gravity should be twice the value expected from Newtonian gravity at roughly 1.75 arcseconds. This was first confirmed by Eddington during his famous expedition to measure a solar eclipse in 1919 \parencite{1920RSPTA.220..291D}.  
    \item Gravitational redshift of light: This test does not have a Newtonian analogue. The basic idea is that the gravitational redshift of light will result from the fact that there is a difference in proper time depending on where observers are in a gravitational field. Early attempts offered some measurements of this effect in the spectral lines of stars \parencite{Wheeler}; however, gravitational redshift was more conclusively demonstrated by the Pound-Rebka experiments \parencite{Pound1959, Pound:1960zz}.
    %can be seen from considering two stationary observers who are at different locations in a gravitational field. One observes the emission of the photon and the other observes the emitted photon far away from the first observer. There is a difference in proper time between the observers who are at different $r$, and consequently, they will measure different frequencies for the emitted photons depending on where they are in the gravitational field. While early attempts were made to measure this effect in the spectral lines of stars, gravitational redshift was only conclusively demonstrated by the famous Pound-Rebka experiments in 1959. 
\end{enumerate}

As both \textcite{Wheeler} and \textcite{Dicke1957} noted at the famous 1957 Chapel Hill conference, at that point in time, the experimental evidence used to infer support for GR was not significantly better than it was only a few years after the advent of the theory. Essentially, the only significant work that had been done in this area involved incrementally improved versions of these classic tests. \textcite{Peebles:2016ahh} attributes the lack of attention afforded by physicists in the first half of the 20\textsuperscript{th} Century to the further testing GR to a number of factors, including nuclear and particle physics consuming most of the oxygen within the community and the seeming absence of any technologically feasible alternative experiments. However, he also attributes this neglect the community's unreserved, overwhelming acceptance of GR due to its compelling theoretical architecture. Considering that the classic tests were the only firm empirical basis for GR for the better part of half a century, they evidently occupy a special place in the history of gravitational physics.

%While this struck Dicke (1957) as ``really very flimsy evidence on which to hang a theory"

The standard view within physics is that these tests (and indeed all empirical tests of gravitational physics to date)  support definitively the idea that gravity is described by a metric encoding spacetime curvature as described by GR.\footnote{In the case of gravitational redshift, the consensus is questioned by \textcite{BrownRead2016}, who argue that the results of Pound-Rebka-type experiments can be accounted for by considering accelerating frames in special relativity. We won't go into these arguments further in this article; see \textcite{FankhauserRead} for further discussion.} For example, \textcite[Sect.~2]{Will},
in one of his many comprehensive reviews of the experimental status of gravitational physics,
%motivates the `Einstein equivalence principle' (EEP) and notes that the EEP ``is the heart and soul of gravitational theory, for it is possible to argue convincingly that if EEP is valid, then gravitation must be a `curved spacetime'".
%Stated simply, the EEP identifies gravity with inertia,\todo{This is how Lehmkuhl states the EEP, but is it really how Will states it? I thought he said something else---worth checking.} which implies that an observer cannot distinguish between being in a state of uniform acceleration or being subjected to a uniform gravitational field.\footnote{For more on EEP, see \textcite{ReadTeh, LehmkuhlRoutledge}. Strictly speaking, the EEP is ambigous between (a) reducing inertia to gravity, (b) reducing gravity to inertia, (c) a symmetric unification of the two (for more on this, see \textcite{Lehmkuhl2014}). Which of (a)-(c) is endorsed will not matter for our purposes in this article.} He then casts the gravitational redshift tests as testing the EEP and the perihelion and light bending tests as testing parameters that govern the degree of spatial and temporal curvature present in the spacetime we are exploring. Thus,
%Will therefore
clearly articulates the conventional wisdom that gravity must essentially be a manifestation of spacetime curvature, with any deviations from pure GR principles and predictions being at best incredibly small.\footnote{For the classic tests, this is usually cashed out from within the framework of the parameterised post-Newtonian (`PPN') formalism \parencite{Will}. In essence, the PPN formalism represents an expansion of the GR metric in terms of its Newtonian approximation and higher order terms that capture the GR effects from the spatial and temporal parts of the metric. This also provides a convenient formalism within which to facilitate comparisons with empirically inequivalent theories of gravity through the behavior of certain higher-order terms in the expansion.} The upshot of this standard position is that modern discussions of empirical tests of gravity almost invariably follow a similar pattern: (1) emphasize that these tests are both compatible with and offer compelling support for GR and its associated spacetime ontology, and (2) explore ways of modifying the standard GR architecture by introducing new fields, degrees of freedom, couplings, etc.,~and studying how the available tests constrain these kinds of modifications (see e.g. \textcite{Workman:2022ynf, Berti:2015itd, Ishak:2018his, Will, Will:2018bme} for some prominent reviews of experimental gravity in a similar vein). Similarly, in one of the most influential and well-cited reviews of modified gravity, \textcite[Sect.\ 2]{Clifton:2011jh} the authors recap the conventional wisdom regarding the foundations of GR as a theory of spacetime curvature, emphasize that all current experimental results are compatible with and support GR as the foundational theory of spacetime, and take the stance that a modified theory of gravity is one for which the field equations ``are anything other
than Einstein’s equations'' \parencite[p.\ 9]{Clifton:2011jh}; before then proceeding to explore systematically how one can modify these equations to produce empirically distinct yet still viable gravitational theories. In what follows, we will examine closely how GR and other theories of gravity pass these tests successfully; in the process,  we will highlight what these empirical studies \textit{actually} test in our gravitational theories.
%\todo{I wonder whether it's worth coming back to Will in the conclusion of our article, and explaining why we don't think he's quite correct here.}

\section{Projective and Conformal Structure in Gravitational Theory}\label{s4}

\noindent Standard foundational presentations of general relativity typically proceed by specifying that kinematical possibilities of the theory are given by tuples $\langle M , g, \Phi \rangle$, where $M$ is a four-dimensional differentiable manifold, $g$ is a Lorentzian metric field on $M$, and $\Phi$ a placeholder for material fields. At the level of dynamics, these fields are specified to satisfy the Einstein equation plus any dynamical equations for the material fields (e.g.,~Maxwell's equations); moreover, typically textbooks make various interpretative stipulations---e.g.,~that metric distances and times are read off by ideal rods and clocks, respectively: see e.g.~\textcite[p.~136]{Malament}.\footnote{\label{fn-causal-inertial}`Causal-inertial' constructivism \emph{\`{a} la} \textcite{EPS} offers an alternative to this `chronometric' approach to the empirical interpretation of GR: see \textcite{LR, ALR2} for discussion. The approach of \textcite{EPS} is by no means devoid of conceptual difficulties---for example, using projective and conformal structures to fix a Lorentzian metric \emph{up to a constant} is still insufficient to write down various GR actions, especially once matter coupling terms are included (our thanks to Brain Pitts for raising this point to us). For further discussion of related issues, see again the above-cited articles; in any case, these issues are by-the-by for our purposes in this article, since we are using the machinery of projective and conformal structures only to understand in a systematic way how GR can be modified without altering its empirical content.}

%There are a number of ways to characterize the structure of geometric theories of gravity. The chronogeometric approach takes the metric to be the fundamental object of interest and postulates the existence of clocks that measure the distance between spacetime events on a particular world line (Synge, 1960). The advantage of this approach is that such clocks are measured locally and the metric $g_{\mu\nu}$ can be used to derive any other quantities of interest. The semantic approach characterizes a gravitational theory in terms of sets of models, where these models are defined by tuples of the form $\left<O_i, ... O_n\right>$. These $O_i$ are mathematical objects, e.g.~tensor fields on a differentiable manifold. For example, a common characterization of general relativity holds that it is a theory given by sets of models of the form $\left<M, g_{\mu\nu}, \Phi\right>$ that also obey the dynamical Einstein field equations (Pooley, 2013). Here, $M$ is a differentiable pseudo-Riemannian manifold, $g_{\mu\nu}$ is of course the metric, and $\Phi$ are matter fields. The advantage of this approach is that one clearly specifies the primary mathematical models of interest that figure in the construction of models from the theory as well as the dynamics that these objects obey. Yet another approach, and the one that we will adopt here, formulates geometric theories of gravity in terms of their projective and conformal structures (Ehlers et al).

In this article, we proceed somewhat differently, by treating the basic geometric object(s) of the theory not as a Lorentzian metric field $g$, but rather as projective structure $\mathcal{P}$, defined as the equivalence class of affine connections $\Gamma$
\[\Gamma \overset{\mathcal{P}}{\equiv} \Gamma' \quad \Leftrightarrow \quad \text{$\exists$ a 1-form $\psi$ s.t.~${\Gamma'}\indices{^{\mu}_{\nu\rho}} = \Gamma\indices{^{\mu}_{\nu\rho}}+\delta\indices{^{\mu}_{\nu}}\psi_\rho + \delta\indices{^{\mu}_{\rho}}\psi_\nu$,}\]
and conformal structure $\mathcal{C}$, defined as the equivalence class of Lorentzian metrics $g$
\[g \overset{\mathcal{C}}{\equiv} g' \quad \Leftrightarrow \quad \text{$\exists$ a function $f$ on $M$ s.t.~$g' = e^f g$} \]
---such a proposal has also been made by \emph{inter alios} \textcite{Stachel:2010zza}. Since \textcite{Weyl1921}, it has been known that a Lorentzian metric is fixed uniquely by its associated projective and conformal structures; since the seminal paper of \textcite{EPS}, it has also been known that a given $\mathcal{P}$ and $\mathcal{C}$, subject to some additional (supposedly innocuous) conditions, fix uniquely a Lorentzian metric.
%(\textcite{EPS}, therefore, can be understood as proving an existence, rather than uniqueness theorem.)\footnote{For further critical engagement with EPS, see \textcite{LR, ALR2}.}

The merit of working with the sub-metrical constituents $\mathcal{P}$ and $\mathcal{C}$ of a Lorentzian metric $g$ is that doing so affords one greater flexibility to explore the results of modifying such structures in turn (a point also made by \textcite{Stachel:2010zza}). Before we get to this, however, we should recall the canonical views on the physical significance of $\mathcal{P}$ and $\mathcal{C}$. Projective structure  $\mathcal{P}$ identifies certain trajectories as being geodesics; empirically, it is supposed to be picked out by the paths of unforced test particles. Conformal structure, on the other hand, refers to the distribution of light cones, and is supposed to be identified empirically by the paths of light rays. (For further discussion of these empirical considerations, see again \parencite{EPS}.)
%refers to the geodesics that freely falling particles follow within such a theory;
%and is determined by the affine connection. Conformal structure $\mathcal{C}$refers to the distribution of light cones and is determined by the metric. A further motivation for this approach is that spacetime theories require matter-energy input in order to describe the world. Massive particles probe projective structure as this structure is closely associated with particles traveling in timelike paths. One the other hand, massless particles probe conformal structure as this structure singles out the class of null-hypersurfaces. In other words,
As \textcite{Stachel:2010zza} writes, ``the relation between conformal and projective structures reflects—and is reflected by—the relation between classical massless wave theories, which in practice means electromagnetism, and classical particle theories and their ensembles represented by the stress-energy tensors of ordinary matter."

\begin{comment}
\begin{itemize}
    \item What are the basic mathematical objects?
    \item How do these mathematical objects have operational significance?
    \item Is the kinematics/dynamics distinction fundamental? (\textbf{James to do})
    \item We could also adopt this way of doing it: define a spacetime theory as $\left(M, g, \Gamma\right)$ where $M$ is manifold, $g$ is conformal structure, and $\Gamma$ is projective structure. 
\end{itemize}
\end{comment}

%More technically, projective structure $\mathcal{P}$ refers to an equivalence class of affine connections [$\Gamma$]. These affine connections determine geodesics that are equivalent up to a reparametrisation (Malament, 2012, pg. 59-60). Consider how an affine connection transforms under projective transformations $\Gamma_{\mu\nu}^{\kappa}{'} \rightarrow \Gamma_{\mu\nu}^{\kappa} + \delta^\kappa_\mu \psi_\nu + \delta^\kappa_\nu \psi_\mu$, where $\psi$ is a 1-form.
In defining projective structure, it is helpful to find an object which is invariant under projective transformations. This can be found by realizing that the trace of the affine connection $\Gamma\indices{^{\kappa}_{\mu\kappa}}$ transforms as $\Gamma_{\mu}' \rightarrow \Gamma_{\mu} + \psi_\mu$. Thus, we can construct an object $P\indices{^{\kappa}_{\mu\nu}}$:
\begin{equation}\label{projective-connection}
    P\indices{^{\kappa}_{\mu\nu}} =\Gamma\indices{^{\kappa}_{\mu \nu}}-\frac{1}{5}\left(\delta\indices{^\kappa_\mu} \Gamma_\nu+\delta\indices{^\kappa_\nu} \Gamma_\mu\right),
\end{equation}
where this object is invariant under projective transformations. We can then say that when  $P(\Gamma) = P (\Gamma ')$ (indices suppressed), both connections are projectively equivalent and belong to the equivalence class $[\Gamma]$, which is identical to $\mathcal{P}$.

In the context of Lorentzian geometries, conformal structure $\mathcal{C}$ refers to an equivalence class of metric tensors [$g$]. This equivalence class of metric tensors determines the light-cone structure and leaves this structure invariant. This essentially amounts to singling out classes of null-hypersurfaces. For example, $g'$ and $g$ belong to the same equivalence class [$g$] if there is a transformation $g' \rightarrow \Omega^2 g$, where $\Omega (x)$ is a function of scale, that leaves the distribution of light-cones unchanged. Another way of thinking about this is that such scale transformations alter concepts like lengths and volumes, but leave angles intact. Of course, one might reasonably ask what conformal structure outside of the context of Lorentzian geometries could possibly amount to---we'll return to this question in \S\ref{s6}, when we introduce non-relativistic theories of gravity.

Before we move on from this section, we have one further comment to make. Although it is fairly standard in the physics literature (see e.g.~\textcite{MatveevScholz}) to characterise geometrical structures such as projective and conformal structures by means of equivalence classes \emph{per} the above, we agree with various philosophers of physics who have written in various contexts (see e.g.~\textcite{WEATHERALL201834, Pitts, Pitts3, ReadSR}) that it would be preferable to present such structures \emph{intrinsically}---i.e.,~to use just the right amount of structure from the outset in order to characterise the object in question, and no more.\footnote{Among other things, as Brian Pitts (p.c.)~asks, ``What are the Euler-Lagrange equations from varying an equivalence class anyway?''} \textcite{EPS} themselves recognise this (see \textcite{LR} for discussion), and accordingly use a conformal metric density to represent conformal structure---such an object is also used for the same purposes by e.g.~\textcite{Pitts}. For the present article, it suffices to recognise this point: the particular way in which projective/conformal structure is defined will not much matter for the points which we seek to make.

%For example, consider the conformal transformation $g_{\mu\nu}' \rightarrow \Omega^2 g_{\mu\nu}$, where $\Omega (x) $ is a function of scale that changes the lengths of vectors and spacetime intervals. Conformal transformations effectively rescale the metric such that $dx' \rightarrow \Omega dx$, $dt' \rightarrow  \Omega dt$, which obviously leaves the light-cone structure intact in a relativistic theory as $ds^2 = 0 = g_{\mu\nu}'u^{\mu}u^{\nu} = \Omega^2 g_{\mu\nu}u^{\mu}u^{\nu}$. 

%Thus, $g_{\mu\nu}'$ and $g_{\mu\nu}$ would belong to the same equivalence class [$g_{\mu\nu}$].

%Why characterize spacetimes in this way? There are several different ways of proceeding, including talking about spacetimes in terms of chronogeometry or semantic models.\todo{Not quite sure what these mean --- so the extra sentence would help :)} \textbf{Sentence about different options.} \todoWill{Haha this is just flagging something for us to talk about.} These certainly have their merits and are clearly related to the projective and conformal structures we have discussed. 

\section{Classic Tests and the Geometrical Trinity}\label{s5}

\subsection{General Relativity}

Since a Lorentzian metric field is fixed uniquely (up to a constant factor) by its associated projective and conformal structure, one can rewrite the kinematical possibilities of GR as tuples $\langle M, \mathcal{P}_{\text{GR}}, \mathcal{C}_{\text{GR}}, \Phi \rangle$.\footnote{Now noting but setting aside the concerns raised in footnote \ref{fn-causal-inertial}.}
%GR is a spacetime theory defined by models of the following form $\langle M, [g], [\Gamma ]_{GR}\rangle$,\todo{NB: this isn't quite how we put the models of the theory in the previous section.} where of course
Here, the projective structure $\mathcal{P}_{\text{GR}}$ is associated with the equivalence class of affine connections defined by the Levi-Civita connection $\Gamma$ and the conformal structure $\mathcal{C}_{\text{GR}}$ is associated with the equivalence class of metric tensors $g$ differing by (spacetime-dependent) scale transformations. $\Phi$ is a placeholder for material fields.

The Levi-Civita connection, whose components in a coordinate basis are,
\begin{equation}
    \Gamma\indices{^\mu_{\nu \lambda}}=\frac{1}{2} g^{\mu \rho}\left(\partial_{\lambda}g_{\rho \nu}+\partial_{\nu}g_{\rho \lambda}-\partial_{\rho}g_{\nu \lambda}\right),
\end{equation}
is crucial to the conceptualization of GR as a theory because it is the only affine connection that realizes the unification of gravity and inertia.
%\todo{How do we justify that it is the *only* such connction? I think we could be a bit more explicit about this.}
That is, it is this choice that leads to the conclusion that gravity is \textit{not} a force, but rather is a manifestation of spacetime curvature.
%\todo{Will this be obvious to people? What about connections which differ by reparameterisation?}
% The projective structure of GR is closely related to the aforementioned EEP and it is  
% %Stated simply, the EEP identifies gravity with inertia, which implies that an observer cannot distinguish between being in a state of uniform acceleration or being subjected to a uniform gravitational field.\footnote{For more on EEP, see \textcite{ReadTeh, LehmkuhlRoutledge}. Strictly speaking, the EEP is ambigous between (a) reducing inertia to gravity, (b) reducing gravity to inertia, (c) a symmetric unification of the two (for more on this, see \textcite{Lehmkuhl2014}). Which of (a)-(c) is endorsed will not matter for our purposes in this article.}
% %The principle allows Newtonian gravity to be subsumed by General Relativity because it generalizes special relativity to apply to noninertial frames (ie, accelerated frames). 
% the identification of gravity and inertia present in the EEP that implies that gravity is \textit{not} a force, but rather is a manifestation of spacetime-curvature. The only suitable affine connection to realize this unification is \textit{Levi-Civita} connection $\Gamma_{\nu \lambda}^\mu$:
% \begin{equation}
%     \Gamma_{\nu \lambda}^\mu=\frac{1}{2} g^{\mu \rho}\left(\partial_{\lambda}g_{\rho \nu}+\partial_{\nu}g_{\rho \lambda}-\partial_{\rho}g_{\nu \lambda}\right)
% \end{equation}
To see this, let us briefly consider how this principle manifests itself in the mathematical formalism of GR and its description of gravity. Our first point of contact---and indeed the starting point for conducting many of the actual tests of GR---is the geodesic equation,
\begin{equation}\label{geodesic}
    \frac{d^2 x^\mu}{d \tau^2}+\Gamma\indices{^\mu_{\nu\lambda}} \frac{d x^\nu}{d \tau} \frac{d x^\lambda}{d \tau}=0,
\end{equation}
where $\tau$ parameterizes the curve in terms of proper time, and $x^\mu$ are the coordinates in use. This defines the equations of motion that would apply to a massive body or photon that is experiencing gravity as described by GR.
%and is normally derived from an action principle whereby one finds extremal curves for the spacetime metric tensor $g_{\mu\nu}$. 
%the notion of inertia refers to the idea that unaccelerated, straight-line motion is the natural motion that a massive body will experience. Thus, 
In Newtonian physics, a test body undergoing inertial motion will naturally move on straight lines in Euclidean space unless or until acted upon by a force. \eqref{geodesic} generalizes this notion of inertia, or straight-line motion, to include curved spaces as the affine connection represents how basis vectors change given an arbitrary curved manifold. Thus, we can understand the geodesic equation as unifying gravity and inertia: a body experiencing gravitation is no longer understood to move in a path that deviates from straight lines in Euclidean space due to a gravitational force, but rather is now understood to move in straight lines within a curved spacetime geometry created by gravity.\footnote{More technically, one could say this: since the difference between any two connections is a tensor, if gravity and inertia are unified in the Levi-Civita connection, then they will not be with respect to any other connection, for which the RHS of \eqref{geodesic} will contain an additional tensorial piece, to be interpreted as a gravitational force. We will see more of this below.}

What's important to us to note about the Levi-Civita connection is that it is the unique connection which is (i) torsion free and (ii) metric compatible. These conditions, respectively, are $\Gamma\indices{^{\lambda}_{[\mu \nu]}}=0$ and $\nabla_{\rho}g_{\mu\nu} = 0$ \parencite[Ch.~3.1]{Wald:1984rg}.\footnote{See also \parencite{Schouten}, which also contains classic discussion of projective and conformal structures.}
%Furthermore, these conditions are essential to capturing the ideas that gravity and inertia are unified and that gravity is a manifestation of spacetime curvature. 
%The first condition ensures that curves that parallel transport their own tangent vectors are paths that are extremal lengths of the metric and the second condition ensures that the lengths of vectors do not change during parallel transport.
The first condition ensures that vectors that are parallel transported along each other will form a closed parallelogram and the second condition ensures that the lengths of vectors do not change during parallel transport \parencite{BeltranJimenez:2017tkd}. 
%Furthermore, these conditions are necessary to ensure that the trajectory that a massive body or photon experiences conforms to this generalized notion of inertial motion as straight lines within a curved spacetime geometry. 
This connection defines the projective structure $\mathcal{P}_{\text{GR}}$ of GR; to anticipate, other theories in the geometric trinity will modify this projective structure (so, in light of what we've said above, those theories will be interpreted as force theories of gravity).
The full dynamical content of GR is expressed by the Einstein field equations, which are those equations obtained by varying the following action---the Einstein-Hilbert action---with respect to the metric:
\begin{equation}
    S_{\text{EH}} = \frac{1}{2} \int d^4x \sqrt{g}R; \qquad \delta S_{\text{EH}} = 0 \quad \Longrightarrow \quad
    R_{\mu\nu} - \frac{1}{2}Rg_{\mu\nu} = k T_{\mu\nu}.
\end{equation}
Here $R_{\mu\nu}$ is the Ricci curvature tensor, $R$ is the Ricci scalar curvature, $g_{\mu\nu}$ is the metric, and $T_{\mu\nu}$ is the stress-energy tensor for matter fields. The Ricci curvature tensor $R_{\mu\nu}$ is built out of the coefficients of the Levi-Civita connection, which in turn are built out of the metric $g_{\mu\nu}$ and its derivatives.

%\footnote{In GR the metric is the fundamental object and all of the other objects are built from it. \begin{equation}    R_{\mu \nu}=\partial_{\lambda} \Gamma_{\mu \nu}^{\lambda}-\partial_{\nu} \Gamma_{\mu \lambda}^{\lambda}+\Gamma_{\mu \nu}^{\lambda} \Gamma_{\lambda \sigma}^{\sigma}-\Gamma_{\mu \sigma}^{\lambda} \Gamma_{\nu \lambda}^{\sigma}, \quad\text{and}\quad \Gamma_{\nu \lambda}^\mu=\frac{1}{2} g^{\mu \rho}\left(\partial_{\lambda}g_{\rho \nu}+\partial_{\nu}g_{\rho \lambda}-\partial_{\rho}g_{\nu \lambda}\right).\end{equation}}.

Thus, we see that the LHS of the Einstein field equations is entirely composed of the metric, and first and second derivatives of the metric, and these are then organized into mathematical objects that quantify concepts like curvature. All together, these structures afford substance to the idea that GR geometrises gravity in terms of spacetime curvature. That is, gravity is described by a dynamical metric tensor $g$; this metric defines a curved spacetime that is determined by the distribution of mass-energy content, and motion under the influence of gravity conforms to inertial trajectories within this curved spacetime. 

\subsection{Classic Tests of GR}

%How were these ideas tested and confirmed?
In basic terms, the classic empirical tests of GR require (i) solving the Einstein equation to obtain the metric $g$, (ii) predicting how objects (massive bodies and photons) should behave in such a spacetime environment, and (iii) testing those predictions. The classic tests of GR are:
\begin{enumerate}
    \item perihelion precession of Mercury's orbit,
    \item gravitational deflection of light, and 
    \item gravitational redshift of light.
\end{enumerate}

It turns out that all of these tests are essentially probes of the Schwarzschild metric of GR. The Schwarzschild metric is the solution to the Einstein field equations which describes the gravitational field outside of a spherical mass $M$ with no electric charge or angular momentum. The solution is obtained by solving the field equations for a metric that is spherically symmetric, static, and in vacuum. Considering these properties, it is an excellent candidate for many astrophysical applications, including modelling the trajectories of planets in the gravitational field of the sun and modelling the trajectories of photons in the gravitational fields of the sun or earth. The form of the Schwarzschild metric is
\begin{equation}
ds^2 = -\left(1-\frac{2 G M}{r}\right) c^2 d t^2+\left(1-\frac{2 G M}{r}\right)^{-1} d r^2+r^2 \left(d \theta^2+\sin ^2 \theta d \phi^2\right).
\end{equation}
Once one has the form of the Schwarzschild metric, one can feed this into the geodesic equation in order to calculate how test particles like masses or photons would behave in this system. Upon using symmetries to simplify the set of coupled differential equations and imposing that motion happen in the equatorial plane, one obtains the following \parencite[Ch.~5]{Carroll:2004st}:
% \begin{equation} \label{eom}
%     \left(\frac{dr}{d\phi}\right)^2 +  \left(\frac{1}{L^2}\right) r^4 -  \left(\frac{2GM}{L^2}\right) r^3 + r^2 - 2GMr = \left(\frac{2\epsilon}{L^2}\right) r^4 ,
% \end{equation}
\begin{equation} \label{eom}
    \frac{1}{2}\left(\frac{dr}{d\lambda}\right)^2 + V(r)= \frac{1}{2}E^2,
\end{equation}
which is a simple equation describing the classical energy $E$ of a mass with a kinetic energy and potential energy $V(r)$ as a function of radius $r$. Here $V(r) =\frac{1}{2} \epsilon-\epsilon \frac{G M}{r}+\frac{L^2}{2 r^2}-\frac{G M L^2}{r^3}$, where $L$ refers to angular momentum and $\epsilon = g_{\mu\nu}\frac{dx^{\mu}}{d\tau}\frac{dx^{\nu}}{d\tau}$ is a constant of motion, which is $\epsilon = 1$ for massive particles and $\epsilon = 0$ for photons. 

Following for example \textcite[Ch.~6]{Wald:1984rg} or \textcite[Ch.~5]{Carroll:2004st}, one can then solve these equations in order to carry out the classic tests of GR for massive bodies and photons, respectively. In particular, for massive bodies ($\epsilon = 1$) GR introduces an extra term (beyond the standard Newtonian terms) in the effective radial potential in the two body problem that is proportional to $\sim r^{-3}$, which induces the additional 43 arcseconds of precession for Mercury's perihelion that had yet to be accounted-for. Similarly, one can explore this problem for photon orbits ($\epsilon = 0$) and conclude that they should expect to observe roughly $\sim 1.7$ arcseconds of deflection around the sun. This is twice the expected Newtonian value, which comes from the fact that a Newtonian analysis of light deflection in effect only considers the temporal component of the metric tensor,\footnote{The simplest way to understand the effective Newtonian limit of GR is to consider the Newtonian limit where gravity is weak and speeds are low compared to the speed of light. This procedure identifies the effective (Newtonian) gravitational potential as $\frac{2GM}{r}$, or the second term in the temporal component of the Schwarzschild metric. While this is a good approximation in many contexts, it clearly does not work when considering photon geodesics because here the spatial components of the metric are just as significant as the temporal component that approximates the Newtonian potential.} while leaving out contributions from the spatial components that are present in the above equation. Finally, the gravitational redshift of light can be seen from considering two stationary observers in a Schwarzschild geometry, one who observes the emission of the photon and another who observes the emitted photon far away from the first observer. There is a difference in proper time between the observers who are at different radii $r$; consequently, they will measure different frequencies for the emitted photons depending on where they are in the gravitational field.

%as was first conclusively demonstrated in the Pound-Rebka experiment using a gamma ray emitting sample of Fe (Pound and Rebka, 1959, 1960). 

\subsection{Geometric Trinity}

These tests were all taken to offer spectacular evidence for GR and to confirm that gravity is indeed a manifestation of relativistic spacetime curvature. After all, these tests probe a specific solution for the spacetime metric $g$, which within the framework of GR encodes spacetime curvature. However, does this evidence truly single out GR as the correct theory of gravitation for our universe? In fact, we can consider other theories which do not rely upon notions of spacetime curvature. In particular, we can consider how the `Teleparallel Equivalent to General Relativity' (TEGR) \parencite{Aldrovandi:2013wha, Golonev, Bahamonde:2021gfp} and the `Symmetric Teleparallel Equivalent to General Relativity' (STEGR) \parencite{NesterYo, BeltranJimenez:2017tkd} account for the phenomena associated with the classic tests.

TEGR is a theory in which gravity is a manifestation not of spacetime curvature, but rather of spacetime \textit{torsion}. Curvature can be quantified as the angle by which a vector rotates when it is parallel transported along a closed path, or in other words, how the tangent spaces of the geometry roll along the curve. Likewise, torsion can be understood as the way that tangent spaces twist along the curve and is described mathematically by the torsion tensor $T\indices{^{\mu}_{\nu\lambda}} = \Gamma\indices{^{\mu}_{[\nu \lambda]}}$, which we already know vanishes in GR as one of the conditions that defines the unique Levi-Civita connection. TEGR replaces the zero torsion condition in the connection with a zero curvature condition. This results in what is known as the Weitzenb\"ock connection, for which (i) $R\indices{^{\lambda}_{\mu\nu\sigma}} = 0$ and (ii) $\nabla_{\rho}g_{\mu\nu} = 0$, but $T\indices{^{\mu}_{\nu\lambda}} = \Gamma\indices{^{\mu}_{[\nu \lambda]}} \neq 0$ \parencite{Heisenberg}. Thus, we can define TEGR as a spacetime theory with models of the form $\langle M, \mathcal{P}_{\text{TEGR}}, \mathcal{C}_{\text{GR}} = \mathcal{C}_{\text{TEGR}}, \Phi \rangle$, where $\mathcal{P}_{\text{TEGR}}$ is the projective structure associated with an equivalence class of affine connections containing as an element $\Gamma_{\text{TEGR}}$, which is the Weitzenb\"ock connection (which replaces the Levi-Civita connection of GR).\footnote{TEGR is sometimes formulated in terms of vielbeins rather than a metric as this makes the gauge structure of the theory more apparent---see e.g.~\parencite{Aldrovandi:2013wha}. We will mostly stick to the metric formulation in this article, though for further philosophical discussion of the different formulations of TEGR, see \parencite{RuwardRead}. One reason to prefer the metric formulation of the theory over the vielbein formulation is that the former seems to trade in fewer surplus degrees of freedom.} The theory is given by the action
\begin{equation}
    S_{\text{TEGR}} = -\frac{1}{2} \int d^4x \sqrt{g}T,
\end{equation}
where $T$ is the torsion scalar (in analogy with the Ricci curvature scalar $R$), defined as
\begin{equation}\label{Tdef}
T := - \frac{1}{4} T_{\alpha\mu\nu} T^{\alpha \mu\nu} - \frac{1}{2} T_{\alpha \mu\nu} T^{\mu \alpha \nu} + T_\alpha T^\alpha, 
\end{equation}
where in turn $T_\mu := T\indices{^{\alpha}_{\mu\alpha}}$ is the trace of the torsion tensor \parencite[\S3]{Heisenberg}. Note that, just as for the Ricci scalar $R$, one requires the metric $g_{\mu\nu}$ to raise and lower indices on the torsion tensor $T\indices{^{\mu}_{\nu\alpha}}$ in order to construct the objects appearing in \eqref{Tdef}.\footnote{Even if one works with a vielbein formulation of TEGR, one can still raise and lower indices with $g_{\mu\nu}$, now treating this as a derived object.} That the projective structure of TEGR is different from that of GR is straightforward to compute: using \eqref{projective-connection}, one finds that
\begin{equation}\label{projective-TEGR}
\overset{\text{TEGR}}{P}{}\indices{^{\alpha}_{\mu\nu}} = \overset{\text{LC}}{P}{}\indices{^{\alpha}_{\mu\nu}} + K\indices{^{\alpha}_{\mu\nu}} - \frac{1}{5} \delta\indices{^{\alpha}_{\mu}} K\indices{^{\lambda}_{\nu\lambda}} - \frac{1}{5} \delta\indices{^{\alpha}_{\nu}} K\indices{^{\lambda}_{\mu\lambda}},
\end{equation}
where $K\indices{^{\alpha}_{\mu\nu}}$ is the `contortion tensor', discussed further below.

Similarly, STEGR is a theory in which gravity is a manifestation of spacetime \textit{non-metricity}, and curvature and torsion vanish. Non-metricity can be understood as the geometric effect whereby the act of parallel transporting a vector changes the length of this vector. That is, STEGR obeys the conditions (i) $R\indices{^{\lambda}_{\mu\nu\sigma}} = 0$ and (ii) $T\indices{^{\mu}_{\nu\lambda}} = \Gamma\indices{^{\mu}_{[\nu \lambda]}} = 0$, but $Q_{\rho\mu\nu} := \nabla_{\rho}g_{\mu\nu} \neq 0$ \parencite{Heisenberg}. Thus, we can define STEGR as a spacetime theory with models of the form $\langle M, \mathcal{P}_{\text{STEGR}}, \mathcal{C}_{\text{GR}} = \mathcal{C}_{\text{STEGR}}, \Phi \rangle$, where $\mathcal{P}_{\text{STEGR}}$ is the projective structure associated with an equivalence class of affine connections containing as an element $\Gamma_{\text{STEGR}}$, which is the non-metric connection of the theory (which replaces the Levi-Civita connection of GR). This theory is given by the action
\begin{equation}
    S_{\text{STEGR}} =  -\frac{1}{2}\int d^4x \sqrt{g}Q,
\end{equation}
where $Q$ is the non-metricity scalar (again in analogy with the Ricci curvature scalar $R$ and the torsion scalar $T$), defined as
\begin{equation}\label{Qdef}
    Q := \frac{1}{4} Q_{\alpha \beta \gamma} Q^{\alpha\beta\gamma} - \frac{1}{2} Q_{\alpha\beta\gamma} Q^{\beta\alpha\gamma} - \frac{1}{4} Q_\alpha Q^{\alpha} + \frac{1}{2} Q_\alpha \tilde{Q}^\alpha,
\end{equation}
where $Q_\alpha := Q\indices{_{\alpha \lambda}^{\lambda}}$ and $\tilde{Q}_\alpha := Q\indices{^{\lambda}_{\lambda \alpha}}$ are two independent traces of the non-metricity tensor \parencite[\S4]{Heisenberg}. Note again that, just as for the Ricci scalar $R$ and torsion scalar $T$, one requires the metric $g_{\mu\nu}$ to raise and lower indices on the non-metricity tensor $Q\indices{^{\mu}_{\nu\alpha}}$ in order to construct the objects appearing in \eqref{Qdef}.\footnote{Again, even if one works with a vielbein formulation of STEGR, one can still raise and lower indices with $g_{\mu\nu}$, now treating this as a derived object.} Again, that the projective structure of STEGR is different from that of GR is straightforward to compute: using \eqref{projective-connection}, one finds that
\begin{equation}\label{projective-STEGR}
\overset{\text{STEGR}}{P}{}\indices{^{\alpha}_{\mu\nu}} = \overset{\text{LC}}{P}{}\indices{^{\alpha}_{\mu\nu}} + L\indices{^{\alpha}_{\mu\nu}} - \frac{1}{5} \delta\indices{^{\alpha}_{\mu}} L\indices{^{\lambda}_{\nu\lambda}} - \frac{1}{5} \delta\indices{^{\alpha}_{\nu}} L\indices{^{\lambda}_{\mu\lambda}},
\end{equation}
where now $L\indices{^{\alpha}_{\mu\nu}}$ is the `contortion tensor', again discussed further below.

At first blush, these theories appear to be very different to GR. After all, they are built out of entirely different geometrical structures; they reject firmly a fundamental tenant of GR that gravity is a manifestation of spacetime curvature because both of these theories mandate that spacetime is necessarily flat!
%Consequently, it is obvious that $\mathcal{P}_{GR} \neq \mathcal{P}_{TEGR} \neq \mathcal{P}_{STEGR}$ as they fundamentally differ in their choice of affine connection. Yet,
Nevertheless, we can concisely understand all of these theories in terms of systematically altering the projective structure used in the initial construction of GR. To see this, 
%recall the importance of the Levi-Civita connection in constructing GR and determining the relevant geometric structures and interpretations of those structures. Recall also that the Levi-Civita is not the unique connection. It turns out that 
consider that the most general affine connection has the following decomposition: 
\begin{equation}
    {\Gamma^\mu}_{\nu \lambda} =  \stackrel{\circ}{\Gamma^\mu}_{\nu \lambda} + {K^{\mu}}_{\nu\lambda} ({T^{\mu}}_{\nu\lambda}) + {L^{\mu}}_{\nu\lambda} (Q_{\mu\nu\lambda}),
\end{equation}
where $\stackrel{\circ}{\Gamma^\mu}_{\nu \lambda}$ is now defined as the Levi-Civita connection, and ${K^{\mu}}_{\nu\lambda} ({T^{\mu}}_{\nu\lambda})$ and ${L^{\mu}}_{\nu\lambda} (Q_{\mu\nu\lambda})$ are known as the `contortion' and `distortion' tensors respectively and are functions of their respective torsion and non-metricity tensors \parencite[p.~9]{Ortin:2004ms}. The conditions relating to curvature, torsion, and non-metricity that are applied in constructing all of these gravitational theories are effectively constraining the form of the affine connection used in the particular gravitational theory of interest. That is, $\Gamma_{\text{TEGR}} = \stackrel{\circ}{\Gamma^\mu}_{\nu \lambda} + {K^{\mu}}_{\nu\lambda} ({T^{\mu}}_{\nu\lambda})$ and $\Gamma_{\text{STEGR}} = \stackrel{\circ}{\Gamma^\mu}_{\nu \lambda} + {L^{\mu}}_{\nu\lambda} (Q_{\mu\nu\lambda})$ and the affine connections of these respective theories can be understood as being composed of the initial Levi-Civita part along with a non-vanishing piece from torsion or non-metricity. (From these equations, \eqref{projective-TEGR} and \eqref{projective-STEGR} follow very straightforwardly.)

While the projective structure utilized in GR has clearly been altered, the conformal structure has not as the spacetime metric---which encodes said structure---has not changed.
%Affine connections and metrics are, of course, independent concepts. Both TEGR and STEGR keep the conformal structure of GR intact. In other words, i
In the process of building these theories we have only manipulated properties of the affine connection, while retaining the same notions of timelike, spacelike, and null intervals inherited from the basic relativistic conception of causal structure. So, this family of theories all share the same conformal structure such that $\mathcal{C}_{\text{GR}} = \mathcal{C}_{\text{TEGR}} = \mathcal{C}_{\text{STEGR}}$.
%: one straightforward way to see this is that all three theories retain the same metric $g_{ab}$ (which encodes conformal structure), while differing on the salient affine connection (which encodes projective structure).
%. \textbf{Probably need to say a little more}

It is one thing to construct a gravitational theory that is fundamentally different from GR, but another thing entirely for that theory to account for the same empirical phenomena in an equally satisfactory manner. Can TEGR and STEGR actually achieve this? The above decomposition of the affine connection gives us a common language that not only applies when relating the affine connections used within the respective theories, but which can also be used in order to translate between all of the structures used within these theories. 
%In other words, if one is interested in investigating the dynamical content of these theories and comparing them to each other, then one can use these relationships to write the actions of each theory in terms of the geometrical structures of one of the others. 
For example, we can use these mathematical relationships to write the Riemann curvature tensor of GR in terms of the TEGR or STEGR connections \parencite{Wolf:2023rad, Heisenberg}. Upon taking the necessary contractions and summations, one finds the relationship between the curvature, torsion, and non-metricity scalars:
\begin{equation}
    - R = T + 2\nabla_\alpha T\indices{^\lambda^\alpha_\lambda} = Q +\nabla_\alpha\left(Q\indices{^\alpha_\lambda^\lambda}-Q\indices{^\lambda_\lambda^ \alpha}\right),
\end{equation}
which allows one to see immediately the relationship between the Einstein-Hilbert action of GR and the TEGR and STEGR actions. This procedure yields the following translation between the actions of GR and TEGR \parencite[Ch.~9.2]{Aldrovandi:2013wha}:
\begin{equation}\label{EHTEGR}
    S_{\text{TEGR}} = -\frac{1}{2} \int d^4x \sqrt{g} R - \int d^4x \sqrt{g} \nabla_\mu T_{\alpha}^{\alpha \mu} = \frac{1}{2} \int d^4x \sqrt{g}T.
\end{equation}
In other words, the TEGR action is equivalent to the Einstein-Hilbert action up to a total divergence term. 
The same procedure for STEGR 
%has  $ \stackrel{\circ}{\Gamma^\mu}_{\nu \lambda} = {\Gamma^\mu}_{\nu \lambda} - {L^{\mu}}_{\nu\lambda} (Q_{\mu\nu\lambda})$ and 
yields:
\begin{equation}\label{EHSTEGR1}
    S_{\text{STEGR}} = -\frac{1}{2} \int d^4x \sqrt{g} R - \int d^4x \sqrt{g} \nabla_\mu (Q_{\alpha}^{\alpha \mu} - Q_{\alpha}^{\mu \alpha}) = \frac{1}{2}\int d^4x \sqrt{g}Q.
\end{equation}
Again, the STEGR action is equivalent to the Einstein-Hilbert action up to a total divergence term \parencite{BeltranJimenez:2017tkd}. 

The upshot of this is that, despite the fact that all three theories utilize entirely different geometric structures to describe gravity, \textit{they are all dynamically equivalent to each other}. This follows from the fact that total divergence terms, also known as boundary terms, do not affect the equations of motion when variational procedures are used to construct the dynamics of these theories. This means that all three theories (GR, TEGR, and STEGR) obey the exact same Einstein field equations (albeit written in terms of different geometric quantities)---see \textcite{Heisenberg, Capozziello:2022zzh} for further discussions of these results. The three classic tests of GR test only the trajectories of massive bodies and photons according to the Schwarzschild solution of the Einstein field equations. Consequently, any theory whose dynamics are equivalent to the Einstein field equations will give identical predictions for any such tests. Indeed, TEGR and STEGR reproduce this same Schwarzschild solution for the spherically symmetric static spacetimes that we consider for these tests \parencite{Aldrovandi:2013wha, Adak}.
This challenges the notion that these tests offer confirmation for \emph{specifically} GR, and the view of gravity as a manifestation of \emph{specifically} spacetime curvature. TEGR and STEGR give empirically equivalent descriptions for all dynamical tests, while describing gravitational degrees of freedom as following from torsion or non-metricity in a flat spacetime environment.

To say more on this: consider moving from the geodesic equation in GR to the geodesic equation in one of these alternative theories:
\begin{equation}
    \frac{d^2 x^\mu}{d \tau^2}+\stackrel{\circ}{\Gamma^\mu} _{\nu\lambda}\frac{d x^\nu}{d \tau} \frac{d x^\lambda}{d \tau} = \frac{d^2 x^\mu}{d \tau^2}+({\Gamma^\mu}_{\nu \lambda} - {K^{\mu}}_{\nu\lambda}/{L^{\mu}}_{\nu\lambda})\frac{d x^\nu}{d \tau} \frac{d x^\lambda}{d \tau}=0,
\end{equation}
depending on whether we are working in the framework of TEGR or STEGR. That is, generalized inertial trajectories are given by the two terms involving the second time derivative of the coordinates and the affine connection (as in GR), leaving a third term proportional to the contortion/distortion tensor that is readily interpreted as a force that would direct test masses and photons away from geodesics. The actual empirical trajectories that test masses and photons follow are \textit{equivalent} regardless of which theory we work in because the dynamical equations of motion are identical in each theory. However, in GR these trajectories are interpreted as inertial motion following geodesics in curved spacetime, whereas in TEGR and STEGR these same trajectories are interpreted as non-inertial motion resulting from gravitational forces that direct motion away from geodesics in a flat spacetime.\footnote{ Of TEGR and STEGR, a reviewer has raised to us the following questions: ``If the torsion theory really says there is torsion, and the non-metricity theory really says there is non-metricity, then where is it?  If there is really torsion in the world, where are the broken parallelograms?'' To these questions, we would reply with a question of our own: could one not say the same of curvature in GR? One likewise doesn't `bump into' those vectors which indicate curvature, at least directly. Indeed, it seems to us that there is perfect symmetry between the three `nodes' of the trinity \emph{vis-à-vis} the `direct' detectability of the geometric properties in which they are couched. For further elaboration on this point, see \textcite{RuwardRead}.}

\section{Classic Tests and Non-Relativistic Gravity}\label{s6}

%Show how to decompose connections into projective and conformal parts --- do the LC and W conenctions agree on conformal part but disagree on projective part?

%- geometric trinity
%- widen lightcones (obsers)

\subsection{Newton-Cartan Theory} It is fascinating that there are alternatives to GR which can account for the classic tests using geometric properties different from curvature. As we have seen, this can be understood concisely as altering the projective structure encoded by the Levi-Civita connection, while leaving unmodified the conformal structure of the models of GR. Surely, maintaining this conformal structure would be essential to passing these tests as it was these very tests that distinguished Einstein's relativistic theory from Newton's non-relativistic theory? As we shall see, even this seemingly unassailable statement does not hold. 

It is well-known that non-relativistic theories of gravity can be geometrized in analogy with GR. `Newton-Cartan theory' (NCT) typically refers to a geometric theory closely related to Newtonian gravity (via the well-known Trautman geometrisation/recovery theorems---see \parencite[Ch.~4]{Malament}). As standardly presented, this theory is given by models of the form  $\langle M, \tau, h, \nabla, \Phi \rangle$, where $M$ is a differentiable manifold as usual, $\tau$ is a (degenerate) temporal metric, $h$ is a (degenerate) spatial metric, and $\nabla$ is a Newton-Cartan connection.
%which is assumed to be compatible with both $\tau$ and $h$ \parencite[p.~248]{Malament}. 
Some important features of this theory are as follows:
\begin{enumerate}
\item The temporal and spatial metrics  obey the orthogonality condition $\tau_{\mu}h^{\mu\nu} = 0$.
\item There is a ``compatibility condition", in analogy with metric compatibility in GR, ensuring that the derivative operator associated with the connection is compatible with both metrics such that $\nabla_{\rho} t_{\mu} = 0$ and $\nabla_{\rho} h^{\mu\nu} = 0$.
\item The connection is assumed to have zero torsion (from this it follows that e.g.~$\nabla_{[\mu} \tau_{\lambda]}$ = 0).
\item The equation of motion is the geometrised Poisson equation $R_{\mu\nu} = 4 \pi \rho \tau_{\mu}\tau_{\nu}$, where $R_{\mu\nu}$ is the Ricci curvature.
\end{enumerate}
Taken together, test particles in NCT are understood to follow geodesics in a curved spacetime (where the curvature is located in the temporal metric) in a manner that is empirically equivalent to Newtonian gravity,\footnote{In fact, for full empirical equivalence with standard Newtonian gravity, further curvature conditions must be imposed in NCT---see e.g.~\parencite[p.~268]{Malament}. These further conditions won't be relevant to us in what follows.} but in very close analogy to GR regarding its geometric structure. Essentially, NCT reformulates Newtonian gravity into a differential-geometric language similar to that of GR.\footnote{We say this without yet wishing to imply that the theories are `fully' (i.e.,~interpretationally/physically) equivalent---more on this below.}

\subsection{Type II Newton-Cartan Theory} As should be obvious given its empirical equivalence with standard Newtonian gravity, NCT would never pass the classic tests of GR. However, \emph{qua} geometric, non-relativistic theory of gravity, NCT as presented above is not the only game in town. One recent approach begins with a relativistic theory and uses a careful  $1/c$ expansion in order to isolate a novel non-relativistic spacetime theory.
%This essentially amounts to taking the $c \rightarrow \infty$ limit of a relativistic theory\footnote{To be slightly more rigorous, what is being done is that we are defining $c = \hat{c}/\sigma$, where $\sigma$ is a small dimensionless parameter and expanding around $\sigma = 0$ \parencite[sec. 2]{Obers}}.
%A natural place to start would be to explore the space of non-relativistic theories emerge from relativistic theories like GR.
This is precisely what is done by \textcite{Obers}. These authors decompose the GR metric into temporal and spatial metrics using what they call `pre-non-relativistic variables' as well as define a new `pre-non-relativistic connection',\footnote{To be explicitly clear, the standard formulation of Newton-Cartan theory can also be understood to emerge in the limit of GR when one takes the non-relativistic limit of the Levi-Civita connection \parencite{1976_Kunzle}. Due to technical reasons, this particular limit utilizing the Levi-Civita connection severely restricts the resulting non-relativistic theory to have a closed temporal metric such that $dt=0$ (see \textcite{VandenBleeken:2017rij} for further discussion). In formulating NCTII, the authors instead utilize a different connection they call the `pre-non-relativistic' connection which allows for this condition to be relaxed. As we shall see, when this condition is relaxed, the resulting theories can encode strong gravitational effects previously thought to exclusively be the purview of relativistic theories.} expand these respective objects in order $1/c^2$, rewrite the Einstein-Hilbert Lagrangian using these objects, and find the non-relativistic limit to give an action of the form
\begin{equation}
    S = \int d^4x L (\tau_{\mu}, h^{\mu\nu}, m_{\mu}, \Phi_{\mu\nu}),
\end{equation}
where of course $\tau_{\mu}$ and $h^{\mu\nu}$ refer to the temporal and spatial metrics respectively, while $m_{\mu}$ is known as the mass gauge field (a peculiar feature of Newton-Cartan geometries that ensures that Newton-Cartan geometric structures are invariant under non-relativistic gauge transformations) and $\Phi_{\mu\nu}$ can be understood as a geometric/tensorial generalisation of a Newtonian potential. The action itself is somewhat cumbersome and its exact expression need not concern us here (see \textcite{Obers} for the full expression); suffice it to say here that this defines a new fully non-relativistic spacetime theory dubbed \textit{Type-II Newton-Cartan theory} (NCTII).
%characterized by models of the form $\langle M, t, h, [\Gamma ]_{TT}\rangle$, where $\Gamma_{TT}$ refers to the so-called ``twistless torsional" NC connection.

Consistent with our approach with the previous theories, we will discuss both the projective and conformal structure of NCTII in order to highlight how it differs from the theories in the geometrical trinity. 
%As with the geometrical trinity, there is a relationship between the projective structures of normal GR and NCTII. That is, one begins with what 
These authors begin with what they call the `pre-non-relativistic' connection which decomposes the object into temporal and spatial parts, expand in terms of $c$, and take the zeroth order term in the expansion \parencite[sec 2.3]{Obers}:\footnote{The idea is to %take the Levi-Civita connection
%\beq
%\Gamma^\rho{}_{\mu\nu} = \frac{1}{2}g^{\rho\lambda}\left(\partial_\mu g_{\nu\lambda} + \partial_\nu g_{\mu\lambda} - \partial_{\lambda}g_{\mu\nu}\right),
%\eeq 
%and 
expand the connection in perturbations of $1/c$
\beq
\Gamma^\rho{}_{\mu\nu} = c^2 \Gamma_{(-2)}^\rho{}_{\mu\nu} + \Gamma_{(0)}^\rho{}_{\mu\nu} + \frac{1}{c^2} \Gamma_{(2)}^\rho{}_{\mu\nu}  + O\left(\frac{1}{c^4}\right),
\eeq
obtaining the leading order term.}
\begin{equation}\label{eq:TTNCconnection}
    \Gamma_{\mu \nu}^\rho=-v^\rho \partial_\mu \tau_\nu+\frac{1}{2} h^{\rho \sigma}\left(\partial_\mu h_{\nu \sigma}+\partial_\nu h_{\mu \sigma}-\partial_\sigma h_{\mu \nu}\right) \text {, }
\end{equation}
where $\tau_\mu$ and $h^{\mu\nu}$ refer to temporal and spatial metrics respectively and $v$ is an inverse of the temporal metric, in the sense that $\tau_\mu v^\mu = 1$ (note that this inverse is not unique; likewise, the inverse of $h^{\mu\nu}$ is not unique).
%\todo{Are you sure this statement about $v$ is correct? $t$ doesn't have a unique inverse in virtue of its degeneracy...} \todoWill{No it is not unique but they do seemingly indicate that v is the inverse in that section...Marco would you know about this?} \todo{Yes, $v$ is an inverse in the sense that $t_\mu v^\mu = -1$. Then there is a issue that the inverse is not unique, but we usually say it is an inverse because it obeys an equation of the form $a*a^{-1}=1$.}
Formally, \eqref{eq:TTNCconnection} seems very similar to the NCT connection when written in components. However, the NCT connection has zero torsion, i.e.\ $\partial_{[\mu} \tau_{\lambda]} = 0$.\footnote{Technically, torsion vanishing does not automatically follow from assuming a closed temporal metric $dt=0$. The modern way of thinking about Newton-Cartan theory is as a gauge theory of the Bargmann algebra. Consequently, the general expression for torsion in terms of the exterior derivatives of the gauge field can be found in e.g.\ \textcite{Read:2018ogw, Hartong:2015zia, Bergshoeff:2014uea}, where one sees that imposing $dt=0$ alone does not kill all the torsional degrees of freedom as it is also possible to have spatial and mass torsion. However, the form of the connection has already eliminated these other degrees of freedom, so imposing $dt=0$ on this connection kills the remaining torsional degrees of freedom.} Without imposing this condition by fiat, the connection described in \eqref{eq:TTNCconnection} naturally has torsion. Indeed different versions of Newton-Cartan geometries can be categorized by different torsion conditions; NCTII uses the `twistless torsional' condition $\tau_{[\mu}\partial_\nu \tau_{\rho]}= 0$, which allows for torsion, but ensures that the spacetime admits of a foliation into equal time slices \parencite{Obers}.\footnote{The classification of Newton-Cartan geometries by torsion conditions is the following: 
\begin{description}
\item[Newton-Cartan geometry] No torsion, i.e.~$\partial_{[\mu} \tau_{\nu]} = 0$.
\item[Twistless torsional Newton-Cartan geometry] $\tau_{[\mu}\partial_\nu \tau_{\rho]}= 0$.
\item[Torsional Newton-Cartan geometry] No constraints on $\tau$.
\end{description}
} The projective structure $\mathcal{P}_{\text{NCTII}}$ is then defined by this twistless, torsional connection (see \textcite{Obers} for a detailed discussion of geodesics in NCTII). 

The conformal structure of this theory can be understood by considering the limit taken in the construction of this theory. The slope of the light cone in Lorentzian geometries is $1/c$, and the process of taking this limit can be visualized as `flattening' out the light cone as we are expanding around $c = \infty$ \parencite[\S2.1]{Obers}.\footnote{More technically, $c = \hat{c}/\sqrt{\sigma}$, where $\sigma$ is a small dimensionless parameter which is expanded around 0 \parencite[\S2.1]{Obers}.}
%\todo{Maybe we need to explain this more.} \todoWill{I added a little detail}
This is a notable departure from any of the theories we have considered so far. Once the light cones are `flattened' in this way, there no longer remains a lightcone structure defined by a null interval on a spacetime metric.
%\todo{Not sure I completely understand---can we  be more explicit?} \todoWill{I am thinking about it and not sure how to explain it better...in general there are not a lot of resources and what there is is very abstract.}
Given this, one might wonder what becomes of conformal structure in such spacetimes. We can, however, continue to speak of conformal transformations that preserve the direction of the degenerate temporal and spatial metrics, in close analogy to how conformal transformations with a relativistic metric preserve angles while altering lengths and volumes. Following \textcite{DUVAL2017197}, we can associate the conformal structure of this particular variant of Newton-Cartan theory $\mathcal{C}_{\text{NCTII}}$ with an equivalence class of metrics [$t$] and [$h$], whose directions are independently preserved by scale transformations $\Omega(x)$.\footnote{For some philosophical discussion of conformal transformations in non-relativistic spacetime settings, see \textcite{DewarRead}.}

Bringing this all together, models of NCTII can be represented in the following way: $\langle M, \mathcal{P}_{\text{NCTII}}, \mathcal{C}_{\text{NCTII}}, \Phi \rangle$. Given that both the projective and conformal structures of this theory are so different from both those of GR and those of the other relativistic theories we have considered up to this point, and instead are associated closely with structures of other non-relativistic theories, it would be striking if this theory could pass most or all of the tests that were taken to confirm GR, and to refute conclusively Newton's non-relativistic theory. 
%Despite NCTII and NCT having formally identical equations of motion, given by $R_{\mu\nu} = 4 \pi G \rho t_{\mu}t_{\nu}$, 
In fact, it turns out that the twistless torsional condition in NCTII opens up solutions that were otherwise excluded in the torsionless version NCT. Moreover, it is these solutions that encode the kinds of gravitational effects to which the classic tests are sensitive---phenomena for which the traditional Newtonian theory could not account \parencite{ObersTests}. 

The procedure for determining dynamical trajectories of test particles in NCTII is very similar to that outlined in GR. Solving the equations of motion for the metrics in a background that is static, spherically symmetric, and in a vacuum yields 
\begin{equation}\label{eq:NRsch}
\begin{aligned}
\tau_\mu d x^\mu & =\sqrt{1-\frac{2 G M}{r}} d t, \\
%m_\mu d x^\mu & =0, \\
h_{\mu \nu} d x^\mu d x^\nu & =\left(1-\frac{2 G M}{r}\right)^{-1} d r^2+r^2 d \Omega_2^2. \\
%\Phi_{\mu \nu} d x^\mu d x^\nu & =0 .
\end{aligned}
\end{equation}
While the metrics are degenerate, the above two equations bear an unmistakable resemblance to the standard Schwarzschild solution in GR. Indeed, one can think of the $(1- \frac{2 G M}{r})^{1/2}$ in the temporal component as a lapse function, which is equivalent to the lapse function in a standard ADM 3+1 split in GR. It is noteworthy that the geometry \eqref{eq:NRsch} can also be derived through a $1/c^2$ expansion of the (relativistic) Schwarzschild solution, wherein the Schwarzschild radius $r_s$ is treated as a constant independent of the speed of light. This expansion effectively terminates after one order in $1/c^2$, indicating that the resulting geometry represents a full solution of non-relativistic gravity. Upon utilizing the geodesic equation and imposing motion in the equatorial plane, \textcite{ObersTests} show that this produces results equivalent to those of GR. That is, the resulting set of equations of motion is identical to \eqref{eom} (which is now no longer a surprise given the form of the metric), and this theory produces both perihelion precession and light deflection in accordance with expectations from GR. Furthermore, from the form of the temporal component we can also see immediately that this theory will produce a gravitational redshift equivalent to GR as it reproduces the exact difference in proper time between observers at different radii $r$.

\section{Modern Tests}

\noindent Although it is indeed remarkable that a non-relativistic theory of gravity can yield equivalent descriptions to GR for the empirical tests which were most responsible for ushering in the paradigm change to the latter, one might nevertheless ask at this stage: how far does this underdetermination really stretch, given the preponderance of modern tests of GR? After all, the bleak empirical status of gravitational theory bemoaned by Wheeler and Dicke did not persist for long following the renewed emphasis on the subject (largely due to their efforts). Ever since, there has been a flurry of new empirical tests, driven both by developments in theoretical understanding and technological advancements. However, it turns out that almost none of these tests can discriminate between GR (along with the other geometric trinity theories) and NCTII! While it would be impractical to list all of them and we make no attempt to do so (see \textcite{Will} for the most comprehensive review), the following is illustrative of the main point. 

Among these, there are a number of prominent tests that were designed to discriminate between GR and alternative theories of gravity that are known to be empirically distinct from GR. For example, E\"{o}tv\"{o}s-type experiments are designed to measure the difference between inertial and gravitational mass and Dicke's group improved these results by orders of magnitude \parencite{1964AnPhy..26..442R}, which has continued to this day with satellite experiments \parencite{PhysRevLett.129.121102}. Despite these impressive modern results, experiments of this type were available even in Newton's day because the equivalence of gravitational and intertial masses holds in any Newtonian gravitational theory as well as GR. Similarly, lunar laser ranging experiments have been used to demonstrate that the gravitational constant does not vary in time \parencite{Muller:2007zzb, Merkowitz:2010kka}, a key prediction of many alternative theories of gravity that violate the strong equivalence principle.\footnote{Constraints on the time variations of the gravitational constant also come from indirect and direct tests of gravitational waves. See \textcite{Wolf:2019hun, LISACosmologyWorkingGroup:2019mwx, Manchester:2015mda} for further details.} Again, this is something on which the theories with which we are concerned agree, because none of them invoke features that would lead to the kinds of signals these experiments are meant to probe.

Many of the other prominent tests of GR amount to further tests that depend on the Schwarzschild solution. These include Shapiro time delay \parencite{Shapiro}, gravitational redshift of light outside of a supermassive black hole \parencite{GRAVITY:2018ofz}, perihelion precession of a star orbiting a supermassive black hole \parencite{GRAVITY:2020gka}, and gravitational lensing \parencite{Bartelmann:2016dvf} (with light deflection being the first example of this kind of test). All of these tests use the Schwarzschild solution as a starting point for determining the predictions that the experiments probe and consequently, NCTII will agree with GR and with the geometric trinity more generally. On the other hand, the Schwarzschild solution is not the only GR solution that has undergone rigorous testing. The Friedmann-Lemaitre-Robertson-Walker (FLRW) solution of GR is a foundational cornerstone of modern cosmology and has likewise been studied empirically extensively \parencite{Planck:2018vyg}. Yet, here again as \textcite{Obers} show, NCTII likewise has a solution that reproduces FLRW cosmology in much the same way that it reproduces the results of the Schwarzschild solution. Thus, NCTII and GR are also in agreement here. 

It is only at the most cutting edge of modern tests that NCTII and the geometric trinity diverge. Most prominently, astrophysical systems such as the in-spiral of black holes and/or neutron stars probe the `strong gravity' regime (as opposed to the `weak gravity' regime of local/solar system tests) and require full numerical treatments of the Einstein field equations to model their dynamical behavior \parencite[Sect.~5.1]{Will}. Just as importantly, these kinds of systems have long been of great interest due to the possibility of them producing observable gravitational wave signatures. 

There has been indirect evidence of gravitational waves since observations indicated that the Hulse-Taylor binary was undergoing an orbital decay consistent with predictions from GR that such systems should radiate away gravitational energy \parencite{HulseTaylor1, Taylor}. However, the first direct detection of gravitational waves occurred in 2016 when the LIGO and Virgo collaborations observed the in-spiral of a binary black hole system \parencite{LIGOScientific:2016aoc}. 
%This test essentially probes the Kerr metric (the solution for rotating black holes---believed to be a good description for actual astrophysical black holes)\footnote{We also note that important tests constraining possible deviations from the Kerr solution have also been recently carried out by the Event Horizon Telescope \parencite{EventHorizonTelescope:2020qrl, Psaltis:2020ctj}.} and perturbations on it in the form of gravitational waves. 
The result was a waveform consistent with both the in-spiral and merger of a binary black hole system and the ringdown of the final Kerr black hole (the solution for rotating black holes---believed to be a good description for actual astrophysical black holes). This is one of the most important recent empirical successes in gravitational research and has constrained significantly the kinds of alternatives theories of gravitational that remain viable \parencite{Baker:2017hug}. The behavior of such astrophysical systems, as well as the existence, production, and properties of gravitational waves, are---unsurprising given their dynamical equivalence---significant empirical consequences upon which all theories in the geometric trinity agree \parencite{Soudi:2018dhv, Hohmann:2018xnb, Bamba:2013ooa, BeltranJimenez:2019tme, Abedi}. 

On the other hand, \textcite{Obers} have found that NCTII does not admit gravitational wave solutions due to the nature of the expansion taken in deriving the theory: \emph{in this sense}, NCTII is neither dynamically nor empirically equivalent to GR. This is consistent with the conventional wisdom that Newtonian theories (both standard Newtonian gravity and traditional NCT) do not admit gravitational waves.\footnote{Although it should be noted that the extent to which wave solutions exist in Newtonian theory is not settled conclusively. The main issue is that the standard Poisson equation, on which Newtonian gravity, NCT, and NCTII all agree, is an elliptic equation and these types of equations are not typically understood to have propagating solutions. See \textcite{Linnemann2021-LINOTS-7} and \textcite{Dewar2018-DEWOGE} for alternative interpretations of the Poisson equation, on which it can be understood as admitting propagating solutions.} While it should not be surprising that empirical results have decided conclusively in favour of relativistic theories of gravity over non-relativistic theories, the extent to which the weak underdetermination between GR and NCTII can be pushed is striking. Indeed, this leads to the admittedly bizarre conclusion that the empirical results from experimental gravitational physics have only recently offered conclusive discrimination between relativistic and non-relativistic theories of gravity, given that NCTII passes so many of the other tests (both classic and modern) of GR.

\section{The Geometric Trinity as a Case of Strong Underdetermination}\label{s7}

\noindent Having argued that (a) the theories which constitute the geometric trinity are empirically equivalent, and that therefore (b) if one of them passes any of the myriad tests conceived of in experimental gravitational physics then all three will do so, it seems that there is a good case to be made that the geometric trinity does indeed present a case a strong underdetermination. Granting this in what follows (although see our discussion of \textcite{Knox2011-KNONTA} below), we here survey some of the possible responses to this underdetermination.
%concluding that the discrimination approach is the only plausible (although controversial and inconclusive to say the least) avenue at the moment if one would like to deny that this is a genuine case of strong underdetermination. 

Recall from Section \ref{s2} that when presented with theories which are empirically equivalent, the first question to ask is: are those putatively distinct theories in fact merely different ways of expressing the same theory? That is: are they theoretically equivalent? Since one component of theoretical equivalence is interpretative equivalence, here we must also ask: are they interpretationally equivalent, in the sense that that they postulate the same underlying ontology, make identical claims about the objects which they describe, etc.?
%\todo{Don't we say somewhere else that there is more to theoretical equivalence than interpretational equivalence?}
%This is the standard view of the equivalence of matrix and wave mechanics (cite things and note Muller).
Exploring the structural differences which exist between the theories which constitute the geometric trinity as we have done deflates any attempt to brush off their differences as merely cosmetic, however. All of these theories postulate different mathematical and geometric structures as underlying our description of gravity; moreover, all three theories disagree on the fundamental character of inertial motion, which is succinctly captured in the differences between their projective structures. In this sense, these theories are interpretationally distinct---for further discussion on this point, see \parencite{RuwardRead}.
%\todo{Do we need to engage with Knox 2011?}

Now, it would be remiss of us at this stage not to draw attention to that fact that \textcite{Knox2011-KNONTA} has presented sustained arguments to the effect that there is, in fact, no genuine underdetermination between GR and TEGR (and, although she does not mention STEGR, we imagine that she would see many of her arguments as extending to that theory also\footnote{\label{fn-knox}That said, it's in fact not completely obvious to us that all of Knox's arguments against TEGR would indeed also apply to STEGR. For example, one of these arguments has it that, since the connection coefficients in TEGR have an antisymmetric component, they cannot be made to vanish in normal coordiantes at a point, and so (Knox avers) there is no TEGR-specific standard of inertial motion. This argument would  not apply in the case of STEGR, since the connection coefficients in that theory are symmetric.}). If Knox is correct, then many of the concerns raised by the existence of the geometric trinity in this article are in fact moot. However, in response to \textcite{Knox2011-KNONTA}, \textcite{RuwardRead} have recently argued that Knox's arguments are at least questionable (in particular, for example, they rely on a particular version of her `spacetime functionalism' which not everyone need be inclined to accept). For what it is worth, we find the arguments given by \textcite{RuwardRead} to be compelling---however, we do maintain that even those more positively disposed to Knox's arguments can find value in our article: what we discuss here represents the various plagues and pathologies from which Knox is able to save us!\footnote{ Moreover, \textcite{Knox2011-KNONTA} does not consider NCTII as this theory had not yet been formulated at the time of her article; further, we hope that our classification of theories in terms of projective and conformal structures is illuminating in any case.}

Granting, then, that this first move (i.e.,~that of insisting upon interpretative equivalence) can be called into question, we are left with (at least) four distinct approaches to dealing with cases of strong underdetermination between empirically equivalent yet distinct theories: roughly following the terminology of \textcite{LeBihan2018-LEBDAO}, we call these the `common core', `overarching theory', `discrimination' and `conventionalist' approaches; here, we discuss each in turn in the context of the geometric trinity.\footnote{\textcite{LeBihan2018-LEBDAO} call their fourth class of response `pluralism'; we'll replace this with `conventionalism' (in the sense of \textcite{DurrRead}), since the latter is a little more precisely-defined, which will suit our purposes.}
%the discrimination approach, the Common Core approach, and the Overarching Approach. 

\subsection{Common Core}

The common core approach advocates moving to a new interpretive framework that allows one to break the underdetermination. This would involve isolating the `common core' that is shared between GR, TEGR, and STEGR, and then interpreting this shared common core as a distinct, ontologically viable theory of its own. 

To see how this might work, it might be helpful to recall a previous example of underdetermination and a successful application of this approach. Consider the well-known example of empirically equivalent Newtonian models of the universe which differ by a kinematic shift. Here, one can isolate the shared common core between the Newtonian models, while purging what is not shared between them (i.e.,~trans-temporal identities of spacetime points \emph{qua} spatial points), in order to arrive at the by-now standard Newtonian mechanics set in Galilean spacetime. Furthermore, the remaining structure can be readily interpreted as a Galilean spacetime rather than a traditional Newtonian spacetime. This provides a clear `common core' interpretation of the two models (one at rest and one with a constant velocity) whereby they completely agree about the structure they attribute to the world (on this, see \textcite{Butterfield}).

This strategy seems viable for GR, TEGR, and STEGR, albeit with an important caveat. Here we have approached the differences between the nodes of the geometric trinity in terms of differences in their projective structures.
%Furthermore, we recall that projective structure is associated with an equivalence class of affine connections while the conformal structure is associated with an equivalence class of metric tensors.
At first blush, it might seem that in the process of isolating their common core by purging them of the structures on which they disagree (including projective structure), one would be left with an impoverished ontology. After all, given that all three theories disagree on projective/affine structure and that this very structure encodes gravitational effects, its not clear what structure is left to actually describe gravity once this is expunged. However, we must recall that the affine structure of GR, defined by the Levi-Civita connection, represents the \textit{unique} symmetric, metric-compatible connection. This means that even when working from the ontology of STEGR or TEGR, one can always use this data to construct the affine structure of GR{, because even in TEGR and STEGR one can define a metric and therefrom its Levi-Civita connection. This is not surprising considering that the affine structure in TEGR and STEGR can always be decomposed into the Levi-Civita piece and the piece given by the contortion/distortion components.\footnote{For further discussion of this point, see \textcite{March:2023trv}.}

GR, in other words, \emph{just is}} the common core of the geometric trinity.\footnote{We are grateful to Adam Caulton and Oliver Pooley for discussion on this point.} The common core approach is usually a very sensible way to interpret theories; however, here there are arguably some reservations about a full endorsement. In many instances (including the previous example), the exercised structure is known to be totally superfluous (i.e.,~retaining fundamental trans-temporal identities of spatial locations does not enhance our modelling of the world in any discernible way); however, it is not clear that that is the case in the trinity. The surplus structure here is crucial to some of the reasons why physicists investigate these alternative geometric frameworks for gravity; this `excess' structure encodes additional gauge degrees of freedom that arguably have the potential to bring gravity into closer alignment with the field theory paradigm of particle physics (more on this in \S\ref{discrimination} when we consider the discrimination approach). Consequently, it is not as obvious in this case that the common core strategy is the correct approach because it is not yet clear whether the extra structure the theories disagree on has some important role to play in further theory development. More philosophically, one should also note with \textcite{LeBihan2018-LEBDAO} that identifying a common core to two or more distinct theories does not \emph{per se} eliminate problematic cases of underdetermination; for this, one requires in addition some argument to the effect that the interpretations of the original two theories were somehow pathological.\footnote{When it comes to e.g.~Rosen's bimetric theory (\citeyear{Rosen1}), or M\o{}ller's tetradic formulation of GR (\citeyear{Moller}), one could likewise argue that GR is the `common core'; the same considerations as elaborated in this paragraph would, however, apply in those cases also.}

\subsection{Overarching Theory}

The common core and overarching theory approaches are both similar in that they require moving to a new interpretive framework that allows one to break the underdetermination. Yet, rather than isolating a shared `common core' amongst the theories, the overarching approach would involve a synthesis of the entirety of the theoretic structures contained within GR, TEGR, and STEGR. That is, the overarching approach seeks to embed these theories into a new framework that would allow us interpret them as different facets of same underlying ontology. 

A good example of a successful application of this approach can be witnessed in the equivalence of the Jordan and Einstein `frames' of Brans-Dicke theory \parencite{Lobo2016OnTP, DUERR202110}.\footnote{`Frames' here is not to be understood in the sense of frames of reference. The use of the word `frame' in the context of Brans-Dicke theory refers to different ways of mathematically representing the theory.} Brans-Dicke theory can be cast into two formulations related by a conformal transformation \parencite{BransDicke, Dicke}. One of these is known as the `Jordan frame', in which a scalar field is non-minimally coupled to the Ricci scalar and test matter still follows geodesics of the metric. The other is known as the `Einstein frame', in which the gravitational part of the action takes the usual Einsteinian form but the scalar field exhibits a universal coupling with matter, leading to test masses being forced away from the natural geodesics of this other metric (not coupled to the scalar field). Historically, there was debate regarding which frame was `correct', or whether they were in fact equivalent (see e.g.~\textcite{Weinstein1996-WEISCA, DUERR202110} for some philosophical discussion), but it was later realized that one could understand the equivalence of the Jordan and Einstein frames by moving away from Riemannian geometry and reformulating the theory within the richer framework of an integrable Weyl geometry \parencite{Romero_2012, Lobo2016OnTP}. In  so doing, it becomes apparent that both of these frames---seemingly representing different geometric objects within a Riemannian geometry---are merely different representations
%gauge fixings \todoWill{Is the terminology I used here referring these to gauge fixings confusing?}
of the same invariant geometric objects within the new Weyl geometry.\footnote{In our view, the fact that the Jordan and Einstein frames of Brans-Dicke theory are equivalent under field redefinition does not diminish the interest and importance of demonstrating that both can be embedded into some integrable Weyl geometry.}

While this is no doubt an exciting example of successful application of the overarching theory approach which might \emph{in principle} lend hope to a similar interpretive strategy in the present case, some tempering of this enthusiasm is probably in order, for acknowledging the possibility that there may exist some overarching theory into which GR, TEGR, and STEGR can be embedded is of course very different from actually finding this overarching theory. Indeed, there is no guarantee that such a theory or framework exists actually exists, or \emph{if} it exists that it can be found.\footnote{Perhaps, for two arbitrary theories, one can prove that there invariably exists some overarching theory into which both can be embedded. But this is conjecture; we'll leave putting meat on the bones as a task for future pursuit.} The proof here is in the pudding, and at present we are---to the best of our knowledge---lacking a framework which would be suitable for employing the overarching theory approach.\footnote{One possibility here would be the metric-affine gauge theories of \textcite{HEHL19951}; this, however, requires further investigation.}

\subsection{Discrimination}\label{discrimination}

In the case of the geometric trinity, the discrimination strategy---of preferentially favouring the ontological claims associated with one theory---appears to be both (a) viable, and (b) deployed actively when physicists and philosophers discuss these theories. That said, its application in this particular instance is understandably fraught with controversy and differing opinions. Indeed, there are a number of ways that this approach can be applied, deploying as they do both philosophically- and physically-motivated criteria (and often a mix of both).

For example, \textcite{Knox2014-KNONSS} has argued that (traditional) NCT is the correct spacetime setting for Newtonian physics, because it has less `surplus' structure (by the same kinds of dynamical symmetry considerations as mentioned above in the case of the move to Galilean spacetime) than standard flat-spacetime Newtonian gravity. This position---that we should prefer a theory or framework with less surplus structure---is clearly a substantive philosophical position (albeit currently a popular one: see \textcite{Dasgupta2016-DASSAA-5} for further discussion). As \textcite{Knox2011-KNONTA} subscribes to this position, she argues that we can discriminate between GR and TEGR in favor of GR. This is because TEGR possesses an additional `internal' freedom to perform Lorentz transformations \parencite{ReadThesis}, meaning that TEGR has additional surplus structure when compared to GR.

In addition to endorsing this reasoning, \textcite{Knox2011-KNONTA} furthermore---and separately---maintains that we can discriminate against TEGR because it is parasitic upon GR (so the argument goes) both in terms of its fundamental ontology and its inertial structure. That is, she argues that it is \textit{really} the Levi-Civita connection that is doing all of the physical work in terms of defining conserved quantities as matter follows the minimal coupling rule with the Levi-Civita connection. Furthermore, in accordance with her views on spacetime functionalism (i.e.\ the role of spacetime is played by what defines inertial frames, where inertial frames are those in which the connection coefficients vanish), she argues that the inertial structure of TEGR is just that of GR because it is only the connection coefficients of the Levi-Civita connection that can be made to vanish. Taken together, \textcite[p.\ 274]{Knox2011-KNONTA} discriminates in favour of GR, arguing that the ``only coherent spacetime to be found in these theories (TEGR and GR) is the curved spacetime of GR'' because the ``spacetime'' posited by TEGR is simply the GR spacetime in disguise (upon adopting her metaphysical views of spacetime). Furthermore, given that in STEGR one also decomposes the Levi-Civita connection into a distinct connection plus a correction term (this time in terms of the distortion tensor), Knox's points presented in the case of TEGR carries over straightforwardly to the case of STEGR---thereby, she can maintain that STEGR, like TEGR, is parasitic upon GR.\footnote{Modulo the issues raised in footnote \ref{fn-knox}.}

One can, of course, resist these conclusions. For example, as already mentioned above, \textcite{RuwardRead} argue that neither of the above positions are compelling. One can just as easily cast the conserved quantities/minimal coupling procedure in terms of the fundamental quantities of TEGR---the contortion tensor and the TEGR connection---and equivalently argue that it is \textit{really} the GR structures that are parasitic on the TEGR structures. Furthermore, regarding inertial structure, we can either reject the philosophical position of spacetime functionalism entirely or move to an alternative account of functionalism to that of Knox.
If one is determined to be a spacetime functionalist, it is not clear why the vanishing of connection coefficients should represent the \emph{sine qua non} of spatiotemporality. One could in principle adopt another account of functionalism, e.g.~that of \textcite{Baker2020-BAKKIS-2} based on fundamentality, and argue that TEGR's inertial structure is more fundamental.

%\textcolor{blue}{James anything else?}

%because its inertial structure is in fact still that of GR, because gravitating but otherwise force-free bodies still follow geodesics of the Levi-Civita connection (i.e.\ the paths can either be interpreted as non-geodesic motion of a TEGR/STEGR connection or as geodesic motion of the GR connection, but the actual path is the same regardless). Indeed, given that in STEGR one also decomposes the Levi-Civita connection into a distinct connection plus a correction term (this time in terms of the distortion tensor), Knox's points presented in the case of TEGR carries over straightforwardly to the case of STEGR---thereby, she can maintain that gravitating but otherwise force-free bodies in STEGR still follow geodesics of the Levi-Civita connection, and therefore STEGR, like TEGR, is parasitic upon GR.
%While STEGR is not discussed specifically by Knox, at least one of these arguments would apply to STEGR (surplus structure because the connection is not uniquely determined), while the theory arguably as natural as GR regarding inertial structure\textbf{Lets discuss the inertial structure}.

So, up to this point, we have two motivations which might underlie the discrimination approach in the case of the geometric trinity:
\begin{enumerate}
    \item Fewer surplus degrees of freedom in one theory versus another.
    \item Inertial structure of one theory remaining `physically significant' in the other.
\end{enumerate}
But there are yet further motivations which might underlie the discrimination approach. For example, some physicists who work on TEGR believe that this theory offers non-trivial \emph{benefits} due to its gauge structure. While in this paper we have chosen to present GR, TEGR, and STEGR in their standard formulations (the tradition formulation of GR and the Palatini formulations of TEGR and STEGR) due both to their familiarity and the ease of comparison that this facilitates, all of these theories can alternatively be formulated in terms of vielbeins \parencite{deFelice:1990hu, Aldrovandi:2013wha, Heisenberg}. We will not dwell on the details here, but within the vielbein formulation of TEGR it becomes arguable that TEGR is a gauge theory of the translation group \parencite[Ch.~3]{Aldrovandi:2013wha}.\footnote{At least, so the claim in the physics literature goes---see \textcite{Wallace2015-WALFAB-6} for some push-back.} For some, such reasoning is compelling because if gravity is a gauge theory of spatial translations, this ``explains why gravitation has for source energy-momentum, just the Noether current for those translations.'' \parencite[p.\ 181]{Aldrovandi:2013wha}. Furthermore, preferring this structure can be understood by way of a unificationist perspective:
%\footnote{As \textcite{Aldrovandi:2013wha} acknowledge (and as \textcite{Wallace2015-WALFAB-6} has also registered), TEGR is not quite a standard Yang-Mills gauge theory due to the presence of soldering. We won't go into this further here.}
\begin{quote}
Three of the four known fundamental interactions of nature---namely, the electromagnetic, the weak and the strong interactions---are described [...] as gauge theory. Only [...] general relativity, does not fit in such a gauge scheme. Teleparallel gravity [...] fits perfectly in the gauge template. Its advent, therefore, means that now all four fundamental interactions of nature turn out to be described by one and the same kind of theory \parencite{Aldrovandi:2015wfa}.
\end{quote}

STEGR also has its own adherents and they express similar unificatory motives. In particular, STEGR can be expressed in what is known as the `coincident gauge'. Here, the theory can be expressed in such a way that its action resembles the Einstein $\Gamma\Gamma$ action, but with the conceptual difference that the connection can now (again, the claim goes) be interpreted as encoding a gauge theory of translations \parencite{BeltranJimenez:2017tkd}. This motivates a similar perspective in terms of its closer conceptual unity with the rest of fundamental physics.
%Furthermore, this way of expressing the theory ``exactly reproduces the resummation for a self-interacting massless spin 2 field..."\todo{What do they mean by a `resummation'?} and ``thus provides a more robust relation with the field theory approach to gravity." (ibid) 
Thus, we have the following further motivation which might be taken to underlie the discrimination approach:
\begin{enumerate}
\setcounter{enumi}{2}
    \item One theory better accords with the architecture of the rest of physics than another, in terms of having the same mathematical structure.
    \end{enumerate}
% \todo{Seems to me that this could in fact be cashed out in a number of different ways---let's discuss tomorrow.}

\begin{comment}
In fact though, there's also a fourth possibility here:

\begin{enumerate}
\setcounter{enumi}{3}
    \item One theory meshes with the architecture of the rest of physics better than another, in terms of coupling to other physical theories.
    \end{enumerate}

On the second: you might think that TPG has more coupling opportunities because it uses tetrads, but in fact this isn't obviously true, because (of course) GR also has a vielbein formulation.
    
\end{comment}

We make no claim that (1)-(3) exhaust the kinds of philosphical/physical considerations which might weigh in favour of the discrimination approach.\footnote{In fact, we can think of several other salient considerations right off the bat---but there's little point in extending this list \emph{ad nauseum}.}
In any case, though: very few (if any) physicists take an absolutist tone when discussing these theories and potentially discriminating between them. However, there are principled reasons that factor into their decisions to work within one framework or the other, particularly when it comes to working with formalisms like TEGR or STEGR that are far less popular and understood than the dominant GR paradigm. In this sense, they are using some combination of physically-motivated principles and philosophical positions to discriminate gently in favour of their preferred leg of the trinity.

\subsection{Conventionalism}

Another strategy which might be deployed in the context of apparent case of strong underdetermination raised by the geometric trinity is \emph{geometric conventionalism}. This strategy is explored (both in general and indeed with specific reference to the geometric trinity) by \textcite{DurrRead}, so we will accordingly keep our remarks here somewhat brief. Suffice it to say that geometric conventionalism is a programme on which one simply abstains from assigning truth values to propositions to do with the geometrical degrees of freedom of the theories under consideration (here, the theories which constitute the geometric trinity): much like choosing a coordinate frame rather than another, one can choose one geometric convention rather than another, but no such choice should be afforded `deep' ontological significance. (Note, though, that this is not the same as declaring all propositions to do with geometrical degrees of freedom to be false, which would be more in line with the common core approach.) As articulated by \textcite{DurrRead}, geometric conventionalism could be argued to be an attractive option in the case of the geometric trinity; moreover, insofar as (as discussed above) physicists are generally \emph{not} dogmatic about their preferred geometrical formulation, it is perhaps not unreasonable to place many into the conventionalist camp.\footnote{Is being a conventionalist sufficient to present a `metaphysically perspicuous characterisation' \parencite{TMN, TMNJR} of one's ontological commitments? It's not obvious, but here we'll set aside that question.}
%\todo{Say that physicists can be read as conventionalists}

\begin{comment}
\begin{enumerate}
    \item Straight up interpretational equivalence blocked because these theoriees on their face say different things about the structure of the world (ie, which fields are dynamical objects of interest, gauge structure of gravity, geometric structures etc)
    \item discriminate 
    \begin{enumerate}
        \item Gauge structure of theories - Perreira/Aldrovani
        \item unification
        \item spacetime functionalism
    \end{enumerate}
    \item common core
    \item overarching
\end{enumerate}

\begin{itemize}
    \item What is a good example of discrimination? We should find a good, uncontroversial one in the literature, BUT we would have a good example in this paper if NCTII did turn out to be fully equivalent to GR/TEGR/STEGR because we could very plausibly argue that NCTII is ontologically contingent on GR. It comes from taking a limit of GR so we should priviledge GR's ontology as prior to that of NCTII.
    \item Example 1: Moving to Galilean spacetime is a good example of successful common core. Seems dubious in this case with only common structure being metrical lengths and conformal structure.
    \item Example 2: Einstein/Jordan frames embedded in Weyl geometry successful example of overarching approach. Would need to find this theory. 
    \item How does conventionalism fit into this?
\end{itemize}
\end{comment}

\section{Conclusions}\label{s8}

\noindent There is no doubt that both classic and modern eras of experimental gravitational physics have proven to be enormously successful in terms of their fruitfulness for theory development, confirmation, and elimination. However, despite this success, there are still somewhat surprising gaps in the knowledge that this renaissance of experimental gravitation has left us, which is essentially due to the ability of alternative theories to reproduce exactly the dynamical solutions that serve as the starting point for conducting experimental tests in the first place. However, this is not necessarily a negative, as the most common reactions to instances of underdetermination would indicate. 

%\textbf{In particular, these tests are all dependent on specific solutions that are derived from GR. This is of course not surprising that these solutions are the starting point for experimental tests of gravity, but it is fascinating that these solutions can be exactly produced by a number of theories.}

Consider first the issue of strong underdetermination within the geometric trinity. We take the underdetermination exhibited by the geometric trinity as an invitation for further exploration. We advocate a broadly pluralistic attitude in the absence of strong reasons to pursue any of the articulated responses to strong underdetermination as it seems that these competing theoretical frameworks have something to offer to physicists and philosophers alike. These potentially fruitful avenues include but are not limited to the following:
\begin{enumerate}
    \item Intrinsic conceptual and interpretive clarity with respect to key physical quantities: Both TEGR and STEGR seem to offer conceptual advantages in terms of defining concepts like energy-momentum density. That is, both have the resources to define tensorial energy-momentum densities because, unlike GR, the structure of the theories allows one to separate inertial and gravitational effects \parencite[\S18.2.3]{BeltranJimenez:2017tkd,Aldrovandi:2013wha}. To give another example, TEGR arguably offers a clear conceptual understanding of black hole entropy. Within the TEGR framework, black hole entropy can be expressed as a volume integral rather than in terms of area, which is more consistent with our typical thermodynamic understanding of the concept of entropy \parencite{Hammad:2019oyb}.\footnote{A full foundational analysis of gravitational energy in the geometric trinity will have to wait for another day. In torsionful or non-metric theories, one can indeed represent such an object tensorially; however, the object turns out to have new gauge degrees of freedom which one might regard as being pathological \parencite{Neto:2000ai}. \textcite{Pitts2}, accordingly, likens this to moving a lump in a carpet.}
    \item Calculation facility: Both TEGR and STEGR have actions which include only first derivatives of metric, as opposed to GR, the Einstein-Hilbert action for which includes both first and second derivatives of the metric. This arguably means that TEGR and STEGR are more natural for problems where boundary terms are important as these actions have well-defined variations for Dirichlet boundary conditions (this was explicitly shown for TEGR by \textcite{Oshita:2017nhn} but also applies to STEGR \parencite{BeltranJimenez:2017tkd}). On the contrary, to use GR in the same applications one must supplement GR with the Gibbons-Hawking-York boundary term to remove the problematic second derivatives of the metric by hand \parencite{Gibbons:1976ue}. A particularly interesting application of this feature is that one can use these theories to calculate black hole entropy (which agrees with the GR prediction), but in an arguably more straightforward manner \parencite{Oshita:2017nhn, Heisenberg:2022nvs}. (Of course, one could always use some other GR action---indeed, for the reasons given by \textcite{WolfRead}, we are sanguine about there not being `one true formulation' of GR.)
    \item Extra-theoretic considerations: e.g. quantisation. For example: (a) quantising GR versus TEGR via a path integral might lead to different (empirically significant) instantonic effects (due to the actions of the theories differing by a boundary term); insofar as one \emph{quantised} theory might thereby be preferred on empirical grounds, one might take this to carry over to the classical case (where the theories are empirically equivalent). And (b): if one theory (e.g.~TEGR) can be cast into the form of a Yang-Mills theory (say), and quantising said theories is unproblematic, this might be taken as a reason to prefer the formalism of said theory (again, say TEGR) over another (say GR).
 %   Where these theories have different gauge structures, this potentially suggests different routes towards quantisation \textbf{do you know more about this? I get it in principle but dont know too many details} \color{red} I'm not sure what this means. The only thing that come to my mind is that the gauge-fixing condition for the different theories could be different (i can check about this if needed). \color{black}
    \item Larger theory space: While GR, TEGR, and STEGR are (again) equivalent to each other, in pursuit of cosmological and modified gravity theories that might explain dark energy or offer plausible realizations of inflation it is common to build extensions out of higher order geometric scalars. These classes of theories go by the names of $f(R)$, $f(T)$, and $f(Q)$ gravity respectively, where for example $f(R)$ represents general functions of the curvature tensor and its contractions (\emph{mutatis mutandis} for its torsion and non-metricity counterparts). However, when these theories are extended in this way they are \textit{not} equivalent and this results in a new space of theories to explore in modified gravity and cosmological applications \parencite[\S5.3]{Bahamonde:2021gfp}.
\end{enumerate}

While the above has focused on all the reasons we should be interested in exploring the geometric trinity, what can be said for Newton-Cartan theory in its various formulations? Of course it is interesting that a version of this theory can pass all of the classic tests of gravity and many of the modern ones, but as a non-relativistic theory of gravity how much value does it have besides being of academic interest? It turns out that these non-relativistic geometries have found important applications in physics:
\begin{enumerate}
    \item (Torsional) Newton-Cartan theories constitute essential tools in the study of condensed matter systems, for example by using NCT as a background for modelling the unitary Fermi gas \parencite{Son:2005rv, Bekaert:2011qd} or the quantum Hall effect \parencite{Son:2013rqa, Geracie:2016inm, Wolf:2021ydy}. On the latter: remarkably, it was found that NCT is the correct background on which to model the quantum Hall effect. This is, roughly speaking, because condensed matter systems are non-relativistic and the structure of NCT provides a natural geometric setting that respects the underlying symmetries of the system.
    \item  There is a non-relativistic version of the AdS/CFT correspondence, within which (torsional) Newton-Cartan geometry plays a crucial role. Indeed, in the non-relativistic AdS/CFT correspondence, the bulk spacetime is a so-called Lifshitz spacetime (instead of AdS) while the boundary theory is a CFT on a NCTII background. This correspondence can be used to model many condensed matter systems \parencite{Hartong:2016nyx}.
    \item (Torsional) Newton-Cartan geometry is the cornerstone of two different initiatives in the pursuit of a quantum gravity theory: first, it is fundamental for non-relativistic string theory \parencite{2012a,2017}, in which non-relativistic symmetries assume the same role as Poincaré symmetries for the relativistic string (that is, they are the global symmetries of the string worldsheet). Second, dynamical (T)NC geometries have been shown to be the geometrised version of Ho\v{r}ava-Lifshitz gravity \parencite{2015}, a gravitational theory first proposed by \textcite{2009}.\footnote{It has been suggested that Ho\v{r}ava-Lifshitz gravity can be useful for describing certain cosmological phenomena \parencite{2010}.} Ho\v{r}ava-Lifshitz gravity is power-counting renormalisable, making it possible for it to be quantised canonically.
\end{enumerate}

\begin{comment}

Open problems:

1. Nature of gravitational waves in Type II

2. Fully understanding modern test

3. Understanding non-relativistic geometric trinity

\begin{itemize}
    \item Understand how temporal and spatial metrics interact in NCTII and how this compares to GR. What happens in NCTII for grav redshift experiment if the observer is moving? That is not pure time dilation. 
    \item What are classic tests actually testing? Metric lengths? Spatiotemporal lengths? Spatial lengths? Temporal lengths?
\end{itemize}

\end{comment}

\section*{Acknowledgements}

\noindent We are grateful to Adam Caulton, Enrico Cinti, Patrick D\"{u}rr, Vincenzo Fano, Jelle Hartong, Elearnor March, Brian Pitts, Oliver Pooley, and Jim Weatherall for helpful discussions. We are also grateful to the anonymous referees for very valuable scrutiny. W.W.~acknowledges support from St.~Cross College, Oxford and the British Society for the Philosophy of Science. J.R.~acknowledges the support of the Leverhulme Trust on a Research Fellowship titled `Measuring Spacetime'.

%\noindent [Redacted for anonymous review.]

%\bibliographystyle{dcu}
%\bibliography{refs}
\printbibliography

\end{document}